\documentclass[aps,prx,twocolumn,superscriptaddress,noshowpacs]{revtex4-2}
\usepackage{graphicx}
\usepackage{xcolor}
\usepackage[normalem]{ulem}
\usepackage{float}
\usepackage{bm}
\usepackage{braket}
\usepackage{placeins}
\usepackage{amsmath}
\usepackage{soul}

\newcommand{\new}[1]{{\textcolor{black}{#1}}}

\newcommand{\comment}[1]{}

\begin{document}

\title{Exciton dynamics uncovering electron fractionalization in superconducting cuprates}

\author{A. Singh}
\affiliation{National Synchrotron Radiation Research Center, Hsinchu 30076, Taiwan}

\author{H. Y. Huang}
\affiliation{National Synchrotron Radiation Research Center, Hsinchu 30076, Taiwan}

\author{J. D. Xie}
\affiliation{Department of Electrophysics, National Yang Ming Chiao Tung University, Hsinchu 30093, Taiwan}

\author{J. Okamoto}
\affiliation{National Synchrotron Radiation Research Center, Hsinchu 30076, Taiwan}

\author{C.~T.~Chen}
\affiliation{National Synchrotron Radiation Research Center, Hsinchu 30076, Taiwan}

\author{T.~Watanabe}
\affiliation{Graduate School of Science and Technology, Hirosaki University, Hirosaki, Aomori, 036-8561, Japan}

\author{A.~Fujimori}
\altaffiliation [email: ] {\emph{fujimori@phys.s.u-tokyo.ac.jp}} 
\affiliation{National Synchrotron Radiation Research Center, Hsinchu 30076, Taiwan}
\affiliation{Department of Physics, National Tsing Hua University, Hsinchu 30013, Taiwan}
\affiliation{Department of Physics, University of Tokyo, Bunkyo-ku, Tokyo 113-0033, Japan.}

\author{M. Imada}
\altaffiliation [Corresponding author. email: ] {\emph{imada@g.ecc.u-tokyo.ac.jp}} 
\affiliation{Waseda Research Institute for Science and Engineering, Waseda University, Shinjuku, Tokyo 169-8555, Japan}
\affiliation{Toyota Physical and Chemical Research Institute, Nagakute, Aichi 480-1192, Japan}

\author{D. J. Huang}
\altaffiliation [Corresponding author. email: ] {\emph{djhuang@nsrrc.org.tw}} 
\affiliation{National Synchrotron Radiation Research Center, Hsinchu 30076, Taiwan}
\affiliation{Department of Physics, National Tsing Hua University, Hsinchu 30013, Taiwan}
\affiliation{Department of Electrophysics, National Yang Ming Chiao Tung University, Hsinchu 30093, Taiwan}

\begin{abstract}
{\bf
Electron quasiparticles play a crucial role in simplifying the description of many-body physics in solids with surprising success.  Conventional Landau's Fermi-liquid and quasiparticle theories for high-temperature superconducting cuprates have, however, received skepticism from various angles. A path-breaking framework of electron fractionalization has been established to replace the Fermi-liquid theory for systems that show the fractional quantum Hall effect and the Mott insulating phenomena; whether it captures the essential physics of the pseudogap and superconducting phases of cuprates is still an open issue. Here, we show that excitonic excitation of optimally doped Bi$_2$Sr$_2$CaCu$_2$O$_{8+\delta}$ with energy far above the superconducting-gap energy scale, about 1 eV or even higher, is unusually enhanced by the onset of superconductivity. Our finding proves the involvement of such high-energy excitons in superconductivity. Therefore, the observed enhancement in the spectral weight of excitons imposes a  crucial constraint on theories for the pseudogap and superconducting mechanisms. A simple two-component fermion model which embodies electron fractionalization  in the pseudogap state well explains the change, pointing toward a novel route for understanding the electronic structure of superconducting cuprates.} 


\end{abstract}

\date{\today}

\maketitle

\thispagestyle{empty}


In cuprate 
superconductors above the superconducting transition temperature ($T_c$), various physical quantities show an enigmatic electronic excitation gap called a pseudogap \cite{Imada1998,Keimer2015}. 
In particular, the formation of the pseudogap around the antinodal points of the Brillouin zone manifests itself in the single-particle spectra measured with angle-resolved photoemission spectroscopy (ARPES) \cite{marshall1996, ding1996, loeser1996, lee2007, yang2008, hashimoto2010, hashimoto2014}.   The mechanisms of the pseudogap formation and the superconductivity itself are core issues of high-$T_c$ superconductivity unresolved for decades \cite{Imada1998,Keimer2015,xu2000,yang_rice_zhang2006,Lee_Nagaosa_Wen2006,varma2006,hinkov2008,daou2010,imada2019}. 

To unravel the origin of the pseudogap, likely originating from electron-electron interaction, it is crucial to clarify the two-particle dynamics in a momentum-energy resolved manner. Recent progress in resonant inelastic X-ray scattering (RIXS) spectroscopy has enabled an exploration of such desired particle-hole charge excitation to probe intrinsic exciton dynamics in the presence of the strong electron Coulomb interaction~~\cite{Ament2011}. To understand the pseudogap and the superconductivity, one would also need to take account of the fundamental fact that carrier-doped cuprates do not follow the conventional Fermi-liquid behavior. A possibility of breaking down the quasiparticle picture at the core of the Fermi-liquid theory must hence be seriously pursued to understand the exciton dynamics.

Among the possible breakdowns, the scenario of electron fractionalization is one of the most impactful proposals, drawing attention for critical tests. The concept that an electron, being an elementary particle in vacuum, is splintered into two or more elementary ``particles" in materials is already established in several phenomena in condensed-matter physics. Fascinating examples of electron fractionalization include: Laughlin's solution representing anyon excitation with fractionalized charge in the fractional quantum Hall effect \cite{tsui1982,laughlin1983}, spin and charge solitons in poly-acetylene~\cite{Heeger1988}, and spin-charge separation in a one-dimensional Tomonaga-Luttinger liquid ~\cite{kim1996,kim2006}. The splitting of an energy band into the upper and lower Hubbard bands, denoted by UHB and LHB, respectively, can be regarded as another typical example of electron fractionalization \cite{imada2019}. 

\begin{figure*}[t] 
\centering
\includegraphics[width=1.6\columnwidth]{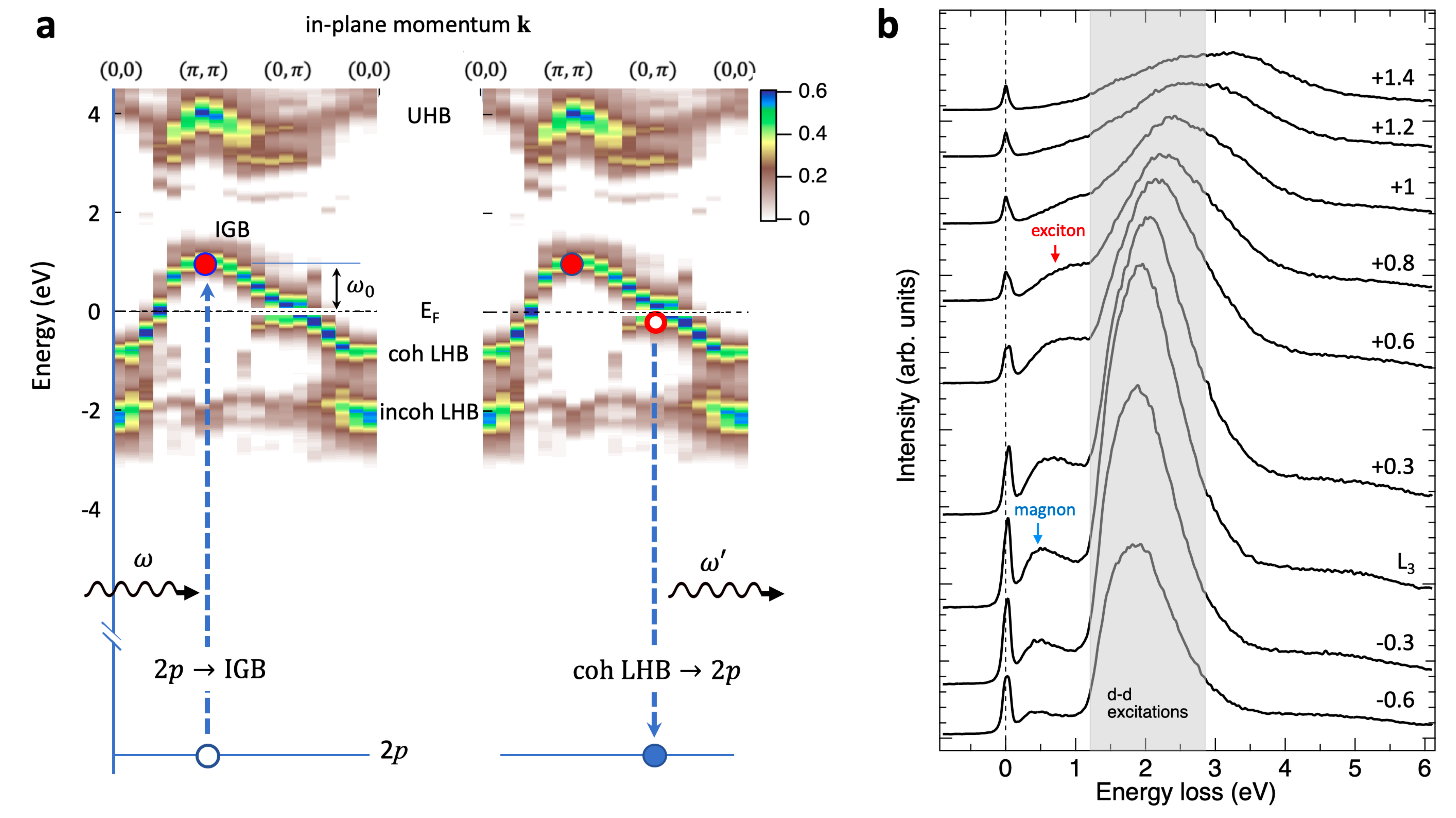}
\caption{\new{{\bf Incident-energy-dependent Cu $L_3$-edge RIXS of hole-doped Bi2212}}. 
{\bf a} Illustration of excitonic excitation induced in RIXS. Left and right panels illustrate electronic transitions of resonant absorption and emission, respectively. Color intensity maps show single-particle spectral functions along a symmetric momentum line reproduced from Fig.~7(a) of Ref.~[28] 
at hole doping 12.5\%. The in-plane momenta ${\bf k}$ are given in units of $1/a$, in which $a$ is the lattice parameter. The coherent part of the lower Hubbard band (LHB) and the incoherent part of the LHB are denoted by ``coh LHB" and ``incoh LHB," respectively. \new{Left: A Cu $2p$ electron is excited to the in-gap band (IGB) near $(\pi, \pi)$ with an energy $\omega_0$ above $E_{\rm F}$. Right: Another electron in the coherent LHB decays to fill the $2p$ core hole through emitting another photon of energy $\omega^\prime$. An exciton is then formed in the combination of red filled circle and red open circle.}
{\bf b}~Cu $L_3$-edge RIXS spectra of ${\bf Q}_{\|} = (\pi, 0)$ for various incident energies across the $L_3$ peak of the XAS. Spectra are plotted with a vertical offset for clarity.  The region in gray indicates the energy loss region of local Cu $dd$ excitation. \new{To identify the exciton peak, we have assumed the following: The energy loss in {\bf b} has one-to-one correspondence to the energy difference between the red filled and open circles in {\bf a} when the momentum difference of the filled and open circles is $(\pi,0)$. The incident energy in {\bf b}  is equal to the red filled circle energy measured from $E_F$ in {\bf a}.} }
\label{Fig.1}\end{figure*}

\new{
Electron fractionalization also embodies the idea of the duality of strongly correlated electrons. The duality is represented by, in one side, the conventional itinerant quasiparticles localized in momentum space connected with the overdoped Fermi liquid. The other side is  electrons localized in real space leading eventually to the Mott insulator in the undoped limit. The duality shares a common concept of the coexistence of itinerant and localized electrons proposed in the literature for metals, for instance in Refs. \cite{Imada1998,kuramotomiyake1990}. As for the localized nature of the electron, as is known in the Mott insulator, an electron is tightly bound to a hole requiring a charge gap to split into an electron and a hole. Such an electron bound to a hole is charge neutral in total and does not primarily interact with an electromagnetic wave. This binding may remain in carrier doped case.}

\new{A two-component fermion model (TCFM), which will be detailed later in this paper, was proposed to embody such a dual character of electrons. The dark fermion called ''$d$ fermion," one of the two constituents of the TCFM represents the localized side of electron bound to a hole induced by doping. The dark fermion is expected to be more stable near the Mott insulator, while the other called ''$c$ fermion" representing the coherent metallic component of electrons becomes more stable in the overdoped region. The TCFM manifests this {\it  bistable} nature of electrons  in lightly carrier-doped systems.}

\new{The idea of electron fractionalization and the TCFM has been studied} to solve the mystery of the pseudogap phase in the high-$T_c$ superconducting cuprates~\cite{sakai2016,imada2019,imada2021} and is supported by the analysis of ARPES data~\cite{yamaji2021}.  
Unlike other possible scenarios involving explicitly symmetry-broken phases with long-range order \cite{xu2000,hinkov2008,daou2010,varma2006}, the distinctive nature of the pseudogap mechanism based on electron fractionalization~\cite{sakai2016,imada2019,imada2021}, as in the resonating valence bond scenario~\cite{yang_rice_zhang2006,Lee_Nagaosa_Wen2006}, does not require a spontaneous symmetry breaking, common to the other types of fractionalization. 

As the information from ARPES
is limited to the dynamics of one particle, analysis combined with two-particle spectroscopy is desired. In the present challenging issue of possible electron fractionalization in the pseudogap phase, such an integrated spectroscopic analysis is particularly important. One can test a recent prediction based on the electron fractionalization that the RIXS intensity is enhanced in the superconducting phase relative to the pseudogap phase~\cite{imada2021rixs}. 

This article presents studies on excitonic excitation of
cuprates using Cu $L_3$-edge and O $K$-edge RIXS. We show an anomalous RIXS enhancement originating from the exciton dynamics in the superconducting phase relative to the pseudogap phase. For this mysterious enhancement, the proposed fractionalization~\cite{imada2021rixs} offers an adequate and consistent understanding. Aside from a unique identification of the origin, our experimental results of the RIXS intensity change establish the atypical exciton dynamics and impose a constraint on theories for the pseudogap and superconducting mechanism.

\vspace{5mm}
\noindent{\bf Excitons revealed by Cu $L_3$-edge RIXS}
\vspace{1mm}

To understand the present RIXS measurement, we discuss the underlying electronic structure calculated with the Hubbard model, which is expected to capture the essential high-energy spectral feature of the cuprates correctly. Figure~\ref{Fig.1}{\bf a} plots the single-particle spectral weight and dispersion of the pseudogap phase calculated with a doped square-lattice Hubbard model~\cite{charlebois2020}. 
The doping of holes into a parent compound of cuprate creates a new low-energy state called in-gap band (IGB) \cite{eskes1991} separated by the pseudogap from the coherent part of the LHB  that crosses the Fermi level $E_{\rm F}$ in and near the nodal region,  as one can see in the blowup shown in supplementary Fig.~S1{\bf b}~\cite{sakai2009}.  The more localized UHB and the incoherent part of the LHB are located at around 4 and $-2$ eV, respectively. 

\begin{figure*}[t!]
\centering 
\centering 
\includegraphics[width=1.8\columnwidth]{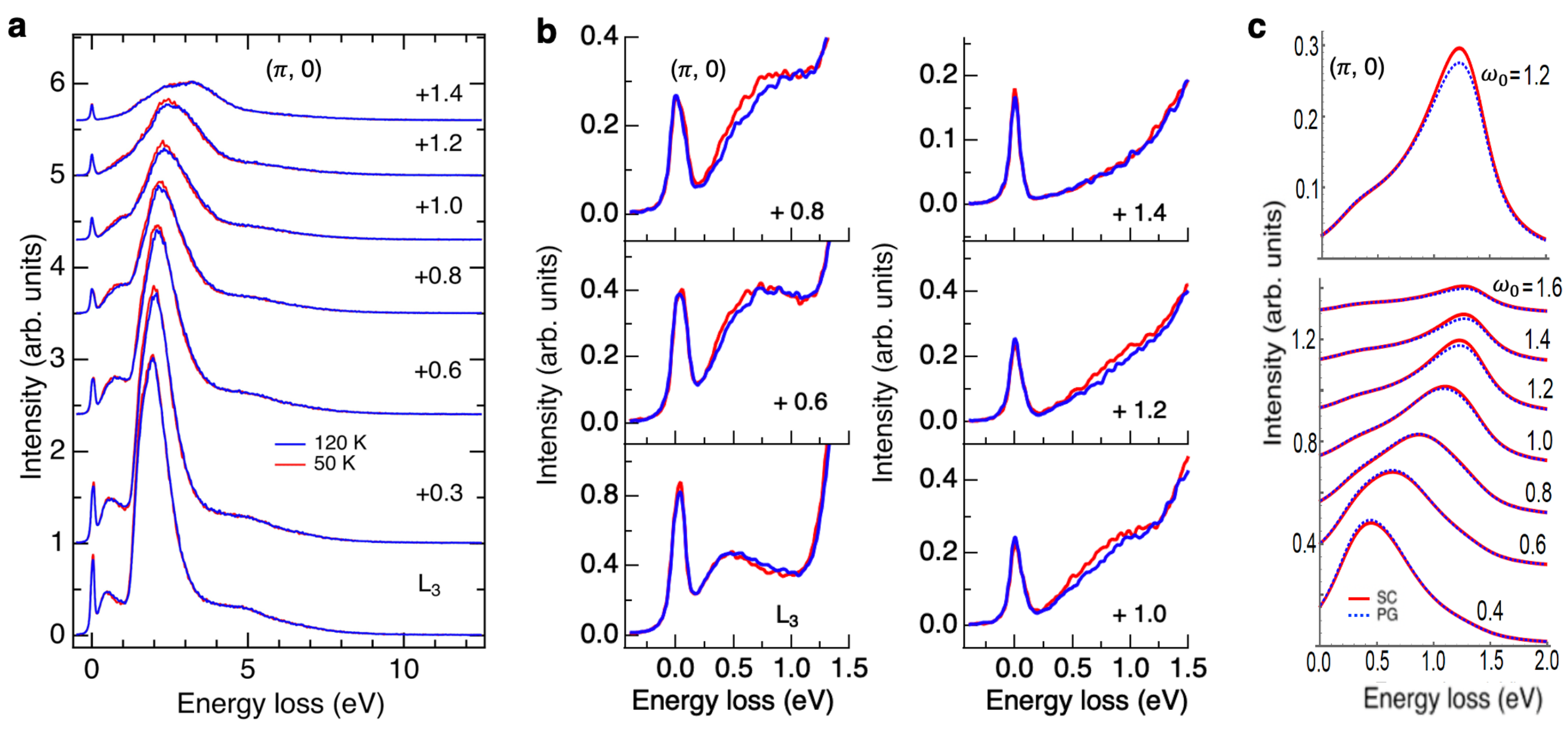} 
\caption{{\bf Excitonic excitation of optimally doped Bi2212}. {\bf a} \& {\bf b}~RIXS spectra at selected incident photon energies for ${\bf Q}_{\|} = (\pi, 0) $  and temperatures above and below $T_c$~=~89~K. The incident photon energy is denoted as its energy relative to the $L_3$ absorption peak in units of eV. Spectra above and below $T_c$ are normalized for energy loss \new{integrated between 1.7 and 13 eV} to highlight the difference caused by the change in temperature; \new{the self-absorption effect of either the incident or the scattered photons was not corrected, because the change in temperature is not affected by the self-absorption}. From the estimate shown in supplementary  Fig.~S3, $L_3$  corresponds approximately to $E_{\rm F}$. {\bf c}~Calculated RIXS resulting from excitonic excitation in the superconducting (SC) (red curves) and pseudogap (PG) phases (blue broken curves) for ${\bf Q}_{\|} = (\pi, 0)$ by using equations (\ref{TCfermionAnomalous}) and (\ref{I_RIXS_1}) as is detailed in Methods and in Supplementary information. \new{The lower panel shows the spectra of selected incident energies $\omega_0$ and the upper panel is the zoom-in spectrum for $\omega_0=1.2$~eV to highlight the enhancement in the superconducting state clearly.} The incident energy $ \omega_0$ measured from $E_{\rm F}$ is given in units of eV.  The core-hole lifetime width $\Gamma$ and the broadening factor $\eta$ were set to 0.3~eV and 0.1~eV, respectively. All spectra are plotted with a vertical offset for clarity.}\label{OP_pi0}
\end{figure*} 

RIXS probes collective excitation as well as excitonic excitation \cite{benjamin2014,guarise2014,minola2015,huang2016,imada2021rixs}. The latter can arise from intra- or inter-band transitions; their excitation depends on the incident photon energy. 
For hole-doped cuprates, the low-energy excitonic excitation results from transitions within the coherent LHB or between the coherent LHB and the IGB. 
Figure~\ref{Fig.1}{\bf b} plots Cu $L_3$-edge RIXS spectra of optimally doped Bi$_{2.1}$Sr$_{1.9}$CaCu$_2$O$_{8+\delta}$ (OP Bi2212) at varied incident photon energies. For the incident energy set to the maximum intensity of X-ray absorption spectrum (XAS), i.e., at $L_3$, we observed distinct Raman-like RIXS features centered about 0.5 and 2~eV. The RIXS feature about 2 eV is predominantly composed of  $dd$ excitation formed by the excitation of a Cu $2p$ core electron to the UHB and then the decay of another $3d$ electron in the LHB to fill the $2p$ core hole. The spectral lineshape of $dd$ excitation reflects the local electronic structure of the Cu $3d$ states; it contains several components derived from transitions determined by the crystal field $10Dq$ and the tetragonal splitting of the $e_g$ and $t_{2g}$ orbitals. 

The RIXS feature of energy loss below 0.5~eV is dominated by magnetic excitation, which becomes most pronounced when the incident X-ray energy is at the $L{_3}$ absorption peak \cite{Braicovich2010,LeTacon2011, Dean2013, Dean2013a, Jia2014, huang2016, minola2015}. \new{As the X-ray energy is increased above $L{_3}+0.6$~eV, this RIXS feature appears to shift toward high energy loss, and evolves into a fluorescence-like excitation. The observed fluorescence-like shift is due primarily to the continuum of particle-hole excitation in the charge channel \cite{Jia2014, huang2016, minola2015}, i.e., an exciton excitation created by the transition of an electron in the Cu $2p$ core level to the IGB (Fig.~\ref{Fig.1}{\bf a}, left panel), followed by the transition of an electron in the coherent LHB to the Cu $2p$ core level (Fig.~\ref{Fig.1}{\bf a}, right panel). The assignment of the peak structure in this energy range to the charge excitation (namely exciton formation) is corroborated by the fact that O $K$-edge spectra shown later also exhibits a similar corresponding peak, indicating that the observed structure arises from the charge channel as the single-spin-flip process is forbidden in the O $K$-edge RIXS.} 

For typical in-plane momentum transfer ${\bf Q}_{\|} = (\pi, 0)$, the exciton is formed by an electron with energy $\omega_0$ above $E_{\rm F}$ near the top of the IGB at $(\pi, \pi)$ and a hole below $E_{\rm F}$ in the coherent LHB near $(0, \pi)$. Depending on the hole energy in the coherent LHB, the momentum transfer ${\bf Q}_{\|}$ does not necessarily bridge precisely between $(\pi, \pi)$ and $(\pi, 0)$, but some intermediate combination can occur in the entire momentum integration over the Brillouin zone. \new{The photon energy $\omega$ determines the energy $\omega_0$ of the excited electron above $E_{\rm F}$, i.e., $\omega = E_{\rm F}+ \omega_0$.} The RIXS energy loss of this exciton is thus pinned by incident X-ray energy $\omega$, because the hole created in the coherent LHB with an energy close to $E_{\rm F}$ has a dominant contribution from the region of the largest density of states near $(0, \pi)$; the energy loss increases with increasing $\omega$.

For a quantitative analysis of the exciton energy, information about the energy $\omega_0$ of an excited electron is essential. The main feature in the Cu $L_3$-edge XAS of hole-doped cuprates originates from the transition into the UHB.  In the RIXS intermediate state, this UHB is pulled down by the $2p$ core-hole potential~\cite{Tsutsui2016}, as illustrated in Fig. S3{\bf a}. 
Similarly, an excited electron in the IGB receives an attractive potential from the core hole to a somewhat smaller extent because of the character of spatially more extended than UHB. This effect leads us to estimate the energy position of $E_{\rm F}$ in the Cu $L_3$-edge absorption spectrum through the fluorescence threshold from an extrapolation; we found that the Fermi level position is near the absorption energy, $E_{\rm F} \sim L_3$. See supplementary Fig. S3.  

\vspace{5mm}
\noindent{\bf \new{Suppression of excitons in the pseudogap phase}}
\vspace{1mm}
 
We proceed to compare the change in the spectral feature of the exciton in RIXS between the superconducting and pseudogap phases. If the temperature is decreased below $T_c$, a superconducting gap~$2\Delta$ $\sim$~80~meV opens in OP Bi2212~\cite{Vishik2012}. The excitonic excitation observed in our RIXS measurements has an energy scale much larger than the superconducting gap. Figure \ref{OP_pi0}{\bf a} displays Cu $L_3$-edge RIXS spectra in a wide range of energy loss with various incident photon energies for ${\bf Q}_{\|}=(\pi, 0)$. Interestingly, Fig. \ref{OP_pi0}{\bf b} shows that the exciton feature of energy loss ${\Delta}E~\sim$~0.8~eV is enhanced in the superconducting phase, particularly for $\omega \sim$~1~eV above $L_3$, which induces a transition from $(0, \pi)$ to $(\pi, \pi)$.  We attribute the difference 0.2~eV between ${\Delta}E$ and $\omega_{0}$ to the exciton binding energy in the final state. A similar enhancement of exciton of ${\bf Q}_{\|}=(\pi/2, 0)$ was also observed, but at a smaller $\omega_0$ ($\sim$~0.8~eV above $L_3$) and with a smaller ${\Delta}E$ ($\sim$~0.6~eV), as plotted in supplementary Fig. S5. The variation in the intensity of elastic scattering, particularly for ${\bf Q}_{\|}=(\pi/2, 0)$, is due to the change in the tail of the scattering intensity from charge density waves, which is beyond the scope of the present paper \cite{Lee2020,HYHuang2021}. 

\begin{figure}[t!]
\centering 
\centering 
\includegraphics[width=1\columnwidth]{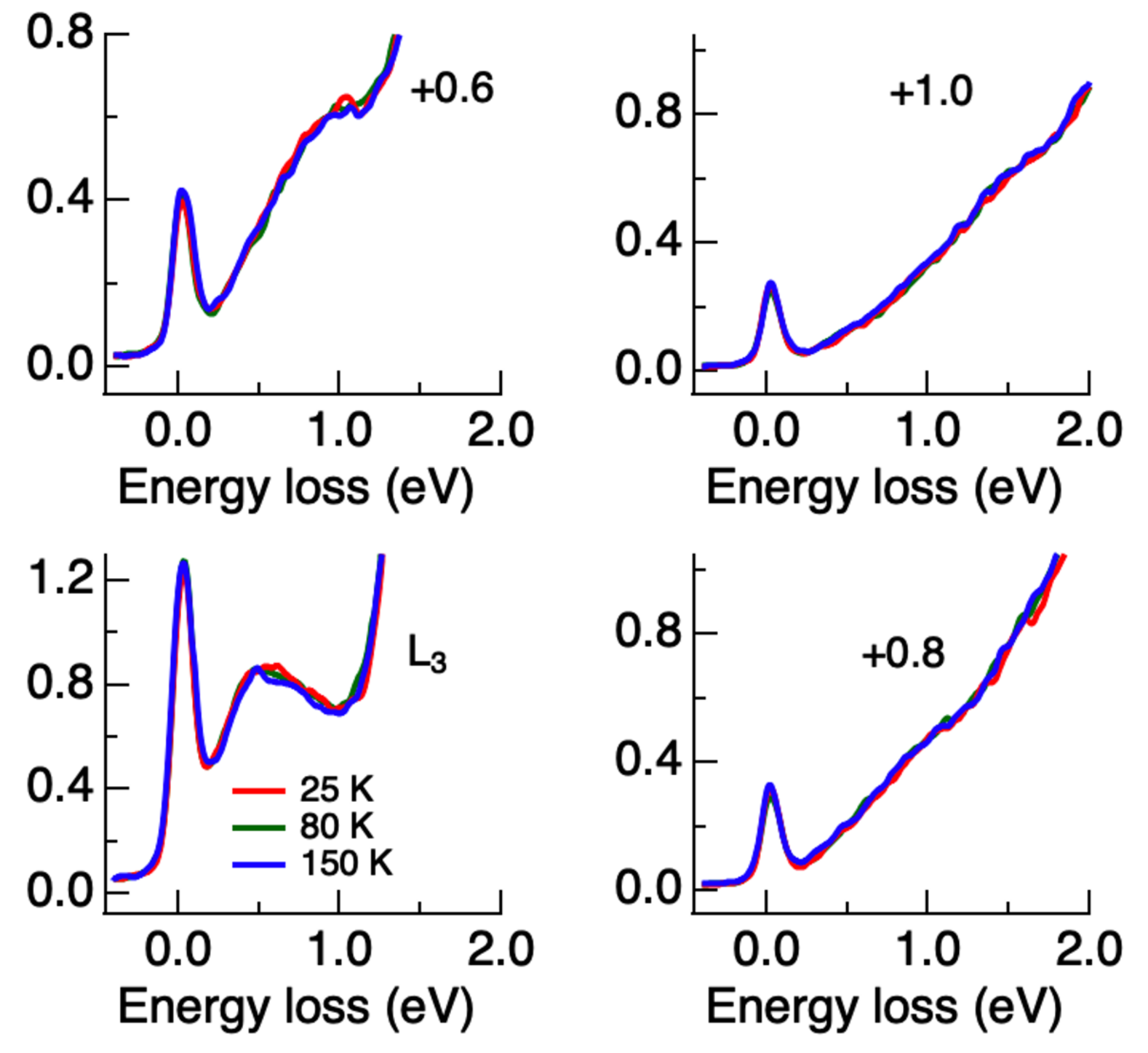} 
\caption{\new{{\bf Excitonic excitation in Cu $L_3$-edge RIXS of overdoped Pb-Bi2212}. RIXS spectra at selected incident photon energies for ${\bf Q}_{\|} = (\pi, 0) $  and temperatures above and below $T_c = 65$ K.  The Pb-Bi2212 single crystal of 
Bi$_{1.6}$Pb$_{0.4}$Sr$_2$CaCu$_2$O$_{8+\delta}$ was with a hole doping level of $p=0.22\pm0.005$. The incident photon energy is denoted as its energy relative to the $L_3$ absorption in units of eV. Spectra above and below $T_c$ are normalized for energy loss from 2 to 13 eV. The self-absorption effect of either the incident or the scattered photons was not corrected. See Fig.~S7 for full-range RIXS spectra}} \label{OD_pi0}
\end{figure}

\new{To gain further insight into the origin of the enhancement in the exciton spectral weight, we resort to measuring RIXS of an overdoped (OD) sample with $T_{\rm c}$ = 65~K, which has no clear signature of pseudogap. Figure \ref{OD_pi0} displays Cu $L_3$-edge RIXS of OD Bi$_{1.6}$Pb$_{0.4}$Sr$_2$CaCu$_2$O$_{8+\delta}$  (Pb-Bi2212) single crystal at temperatures above and below $T_{\rm c}$. 
The shift of exciton energy from the OP to OD samples is about 0.5~eV, consistent with the expected chemical potential shift \cite{Harima2003,Hashimoto2008}. We observed no enhancement induced by the superconductivity in the exciton spectral weight of the OD Pb-Bi2212 sample. This leads us to suggest that the observed exciton enhancement  in the superconducting state (or conversely the exciton suppression in the pseudogap state) in OP Bi2212 is connected to the existence of the pseudogap phase.}

\begin{figure*}[t!]
\centering 
\centering 
\includegraphics[width=1.8\columnwidth]{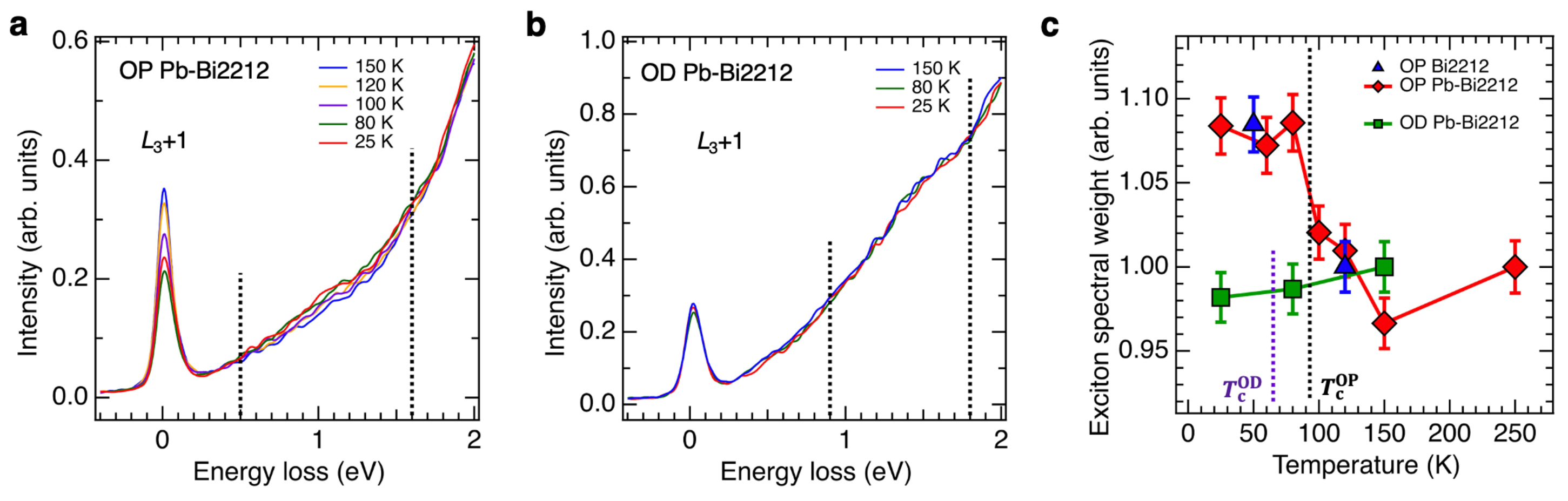} 
\caption{\new{{\bf Evolution of the exciton spectral weight across the superconducting transition.} {\bf a} \& {\bf b} Cu $L_3$-edge RIXS induced by incident photons of energy tuned to 1 eV above the $L_{3}$ absorption energy with ${\bf Q}_{\|} = (\pi, 0)$ for OP and OD Pb-Bi2212, respectively. The normalization scheme of spectra is the same as those described in Figs. 2 and 3. The dotted lines indicate the energy range selected for spectral weight integration. {\bf c} Evolution of the exciton spectral weight after an integration of RIXS spectrum between 0.5 and 1.6 eV for OP and between 0.9 and 1.8 eV for OD samples, respectively, as indicated by dotted lines in {\bf a} and {\bf b}.  The exciton spectral weights of the OP pristine Bi2212 at two temperatures are also included in {\bf c}, as plotted in blue triangles. The vertical dotted lines indicate the transition temperatures $T_c$ of OP and OD Pb-Bi2212 crystals.}} \label{T_dep}
\end{figure*}

\new{Figure~\ref{T_dep} compares the RIXS intensity between the OP and OD samples by examining detailed temperature dependence around the superconducting transition for the RIXS feature arising from the exciton. The temperature evolution plotted in Fig.~\ref{T_dep}\textbf{c} shows a marked contrast between the OP and OD samples. In the OD sample, the intensity shows essentially no temperature dependence.  In contrast, the OP sample shows first a decrease from 250 K to 150 K, where the pseudogap develops. Then further decrease of temperature below the superconducting fluctuation temperature $T_{\rm scf} \sim$~110 K \cite{usui2014} shows a sharp increase of the intensity to the saturated value in the superconducting phase below $T_{\rm c}$.}  

\begin{figure*}[t]
\centering 
\includegraphics[width=1.6\columnwidth]{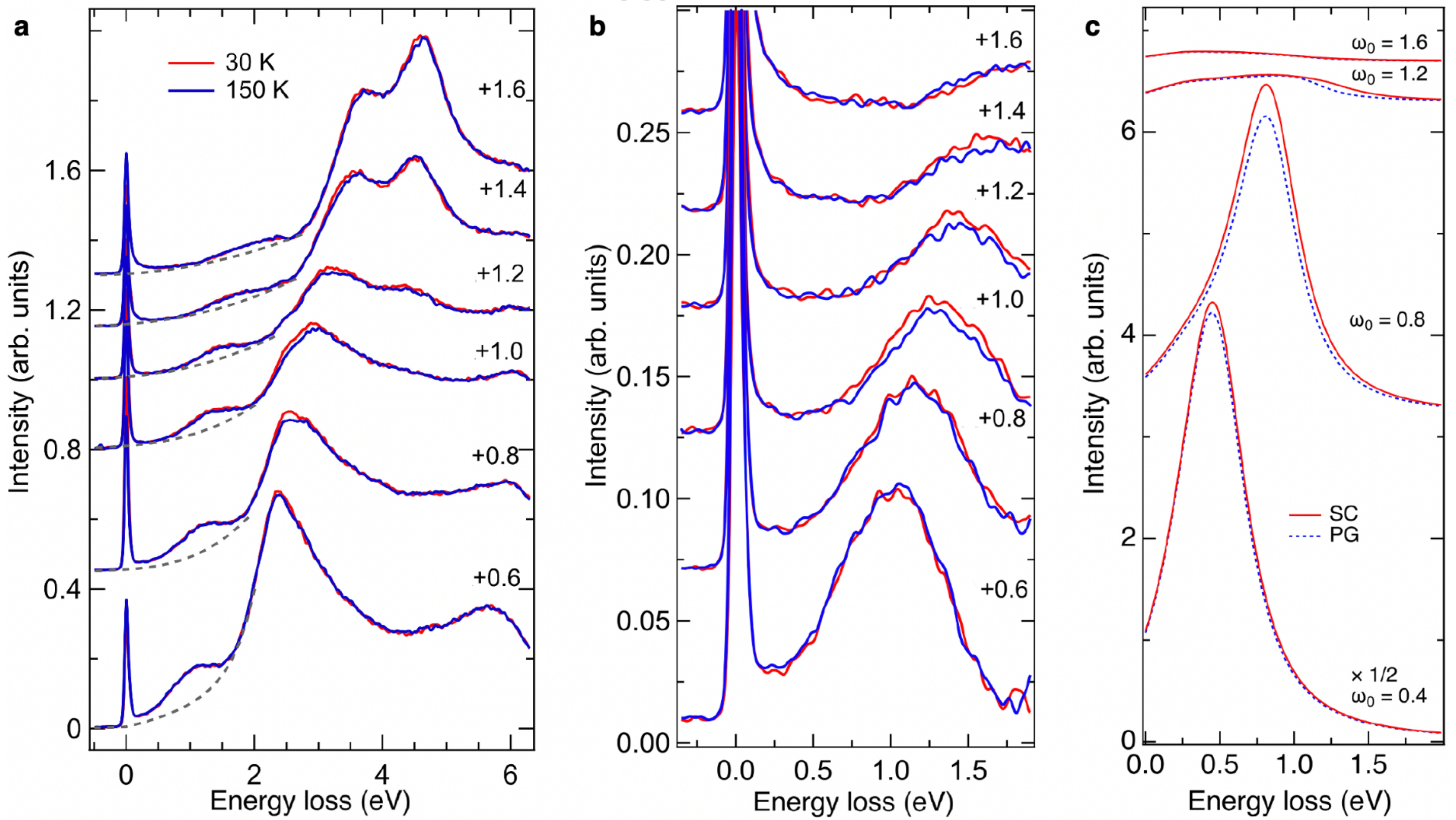} 
\caption{{\bf Exciton excitation in O $K$-edge RIXS of OP Bi2212.} {\bf a}~O $K$-edge RIXS spectra at selected incident photon energies for ${\bf Q}_{\|} = (\pi/2, 0)$ and temperatures above and below $T_c$. The incident photon energy is denoted as its energy relative to the absorption energy $A$ in units of eV. The tail of the RIXS feature created by the transition from the incoherent LHB to the IGB is depicted by a dashed line. {\bf b} Zoom-in spectra of {\bf a} after removing the tail of RIXS induced by the transition from the incoherent LHB to the IGB. {\bf c}~Calculated RIXS for ${\bf Q}_{\|} = (\pi/2, 0)$ with the same condition as in Fig. \ref{OP_pi0}, except for $ \Gamma = 0.07$~ eV and $ \eta  = 0.1$~eV. All spectra are plotted with a vertical offset for clarity.}\label{O_K}
\end{figure*}

Excitons probed with the O $K$ edge also reveal the signature of electron fractionalization. Figures \ref{O_K}{\bf a} and \ref{O_K}{\bf b}  display RIXS spectra of Bi2212  for ${\bf Q}_{\|}=(\pi/2, 0)$ and show the same enhancement across the superconducting transition. In these RIXS spectra, the incident X-ray energy is expressed as the energy of the transition of the $1s$ core electron to the mobile hole band, the so-called Zhang-Rice singlet band (ZRSB). $A$ denotes the energy level corresponding to an absorption energy 528.5 eV. Note that the ZRSB defined in the three-band Hubbard model is  mapped onto the IGB plus coherent LHB in the  framework of the single-band Hubbard model \cite{CCChen2013}.
By analogy with the Cu $L_3$-edge RIXS, we assumed that the position of the bulk Fermi level $E_{\rm F}$ corresponds to $A$ in the O $K$-edge XAS, i.e., $A~\sim E_{\rm F}$. The exciton enhancement in the O $K$-edge RIXS of ${\bf Q}_{\|}=(\pi/2, 0)$ occurs at energy loss 1.3~eV and also between 2.5 and 3~eV with an incident X-ray of energy $A+1$~eV. The 1.3-eV exciton feature is formed by an electron excited to 1~eV above $E_{\rm F}$ and a hole below  $E_{\rm F}$ created by the decay of an electron in the ZRSB near $(0, \pi)$ to the $1s$ core hole.

In conventional understanding, the temperature dependence from 50 to 120 K of the order of 5-10\% in the weight of the RIXS intensity at the 1-eV exciton peak is surprising. \new{In fact, all of the reports in the literature on the exciton measurements, either by RIXS~\cite{suzuki2018,barantani2021} or optical conductivity ~\cite{molegraaf2002,levallois2016}, have proven that the temperature dependence is at most around 1\% of the total spectral weight of the exciton peak. They emphasize either the peak suppression or shift  through $T_c$ by a detailed analysis, but the relative change through $T_c$ is  tiny (roughly $< 1$\%) and it is almost hidden in the background change of the 1\% order or less. This proves that both the background temperature dependence and meaningful change below $T_c$ are very small.} 

We have observed unusual enhancement in the superconducting phase of OP Bi2212, although the temperature change and the exciton excitation energy are similarly two orders of magnitude different. In addition, our RIXS data at the Cu $L_3$- and O $K$-edges show consistent temperature dependence. This is by itself  a surprising result, and must reflect a dramatic change of the electronic structure in that temperature range due to a specific mechanism and cannot be a simple background effect.
\new{Note that the enhancement observed both in Cu $L_3$ and O $K$ edges evidences that it is ascribed to the exciton charge channel and not to the single-spin flip magnetic excitations.}

To the best of our knowledge, the atypical change in the spectral feature described above cannot be explained with a single-component model as plotted in supplementary  Fig.~S6. In this simple single-component case, in proceeding from the normal state to the superconducting state with the BCS mean-field order, the total weight of the quasiparticle component remains unchanged;  the change in the high-energy region should be negligible, although the single-particle spectral weight near $E_{\rm F}$ is decreased because of the superconducting gap formation.
We shall also refer to previous studies that most of the exciton measurements in the literature show essentially no temperature dependence compared to the present conspicuous enhancement below $T_{\rm c}$.

\vspace{5mm}
\noindent{\bf Two-component fermion model analysis}
\vspace{1mm}

The enhancement of excitons at energies 
much higher than the superconducting gap is highly nontrivial.
To understand the origin of this enhancement, we resort to 
the electron fractionalization manifested by the TCFM, because this mechanism offers a satisfactory understanding of the observed enhancement~\cite{imada2021rixs}. 
\new{The electron bistability sketched in the introduction induces the two different tendencies. One is the phase separation into microscopically carrier rich (dominated by $c$ fermion) and poor regions (dominated by $d$ fermion) or inhomogeneity in real space such as the stripe order as is widely observed in the cuprates. The other is a resonant combination of $c$ and $d$ fermions stabilized uniformly in real space as is represented by the TCFM formalism. We focus on the latter. More explicit description of TCFM (Eq.(\ref{TCfermionAnomalous})) and its consequence for the RIXS spectra is found in Methods (See also Supplementary Information for the physical properties of TCFM).}

In this non-interacting model, the hybridized (resonating) $c$ and $d$ fermions generate the bonding  and antibonding bands, which correspond to the coherent LHB and the IGB, respectively. This simple framework well describes the essence of the electron fractionalization~\cite{imada2019}, which enables us to understand the mechanism of the pseudogap of cuprates \cite{sakai2009}. The hybridization occurs in such a way that the fractionalization and resultant pseudogap are most prominent around the antinodal points, i.e., $(\pi, 0)$ and $(0,\pi)$. Although the pseudogap is ascribed to some complex many-body effect in the framework of the one-component interacting fermion system, in the TCFM it is interpreted as a simple one-body hybridization gap of the two constituents, and no spontaneous symmetry breaking is required.

The TCFM model successfully captures the essence of the change in the exciton spectral weight despite the simplification of calculations for RIXS spectra. 
Below we focus on the inter-band transitions which lead to an exciton formed between an IGB particle and a coherent LHB hole, and compare the exciton spectra between the experimental data and the TCFM calculations.
Figures \ref{OP_pi0}{\bf c} and \ref{O_K}{\bf c} plot, respectively, calculated Cu $L_3$-edge and O $K$-edge RIXS spectra for ${\bf Q}_{\|}=(\pi, 0) ~\&~(\pi/2, 0)$ in the pseudogap and superconducting phases using the TCFM. The calculation reproduces the increase of the energy loss with $\omega_0$ in agreement with the RIXS results. The calculations also explain  
the enhancement of the RIXS intensity at around $\omega_0 \sim$1~eV in the superconducting phase. 

The mechanism of the enhancement shown by the TCFM calculation is conceptually explained as follows~\cite{imada2021rixs}. The electronic states in the filled part near the antinodal region in the pseudogap phase contain a large weight of the dark fermion component. However, since the transition to the core hole in the RIXS process is allowed only for the quasiparticle (quasihole) component and not for the dark fermion, a suppression of the RIXS intensity as compared to that above the pseudogap temperature occurs. As the superconductivity is realized by the  Cooper pairing of conventional electrons, the fractionalization is suppressed, resulting in the recovery of quasiparticle component. It leads to an enhanced transition probability to the core hole; in turn, the RIXS intensity increases.

\vspace{5mm}
\noindent{\bf Discussions and Conclusion}
\vspace{1mm}

\new{
Although the experimental and calculated line shapes around 1 eV in Figs.~\ref{OP_pi0} and \ref{O_K} look different at first glance, the difference can be simply attributed to
the strong tail of the well established $dd$ excitation peak at 2-3 eV and other high-energy excitation, and the paramagnon excitation at $<0.5$ eV. After subtracting them out, one can clearly identify the peak structure around 1 eV that can be attributed to the exciton and can be compared with the TCFM calculatoin. Bearing this in mind, we
find remarkable quantitative agreement between the experiment and calculation in the following sense: (a) The exciton peak energy in the RIXS spectra has an overall and quantitative agreement between the measurement and calculation: For instance, at the Cu $L_3$ edge, the peak energy $\omega_{\rm peak}$ is nearly equal to the incident energy $\omega_0$, namely, $\omega_{\rm peak}\sim \omega_0$ in both experiment and calculation at the momentum transfer $(\pi, 0)$. The peak width is roughly 0.5-1.0~eV, which is also in agreement between experiment and calculation. (b) The enhancement in the calculation at the peak energy is 7.6\% for the Cu $L_3$ edge at $\omega_0=$1.2 eV and $\omega_{\rm peak}=1.23$ eV. In the case of experiment, as one sees in Fig.~\ref{T_dep}{\bf c}, the enhancement at temperatures far below $T_c$ is around 7-10\% relative to the pseudogap region around 110~-~150~K.}

\new{In the above comparisons, the parameters in the TCFM model are solely chosen to reproduce the ARPES data without further adjustment. Because ARPES data do not contain the unoccupied part, TCFM parameters may contain larger errors for those related to the unoccupied degrees of freedom.  Nevertheless, we obtained essential agreement of the exciton energy (a) and the enhancement rate (b) between the TCFM calculation and RIXS data. By considering the simplicity of TCFM without parameter adjustment, this quantitative agreement is highly nontrivial. RIXS data contain richer information in the unoccupied part of the single-particle spectra as well as the two-particle exciton spectra. Present basic agreement assures that integrated analyses of ARPES and RIXS together with other spectroscopic data will provide us with more comprehensive understanding in future.  
}

\new{As for the different exciton energies between the O $K$ edge ($\sim 1.3$ eV) and the Cu $L_3$ edge ($\sim 0.8$ eV) with $\omega_0$~=~1~eV, one possible scenario is to take into account the character of the IGB in more detail}. According to a TCFM proposal, the IGB could be a fermionic component of the weakly bound Wannier exciton, in which the excited electron is loosely bound to a nearby hole \cite{imada2019}. The repulsive interaction between the O $1s$ core hole and the bound hole residing away from the excited electron can increase the exciton energy. 
In this interpretation, the IGB state can be regarded as a consequence of the screening of an electron in the unoccupied level by a hole and might bridge the picture of fluorescence in the actual physical process. For the Cu $L_3$ edge, the core hole works instead as an attractive potential for the Wannier exciton, because the Cu core hole dominantly interacts with not the bound hole but the excited electron at the same Cu site, and decreases the energy loss. This effect explains the energy difference of the exciton between Cu $L_3$ and O $K$ edges beyond the present simplified TCFM analysis.

We observed the same enhancement of an exciton excitation with an energy loss centered at about 2~eV, overlapping the tail of the Cu $dd$ excitation; see Fig. \ref{OP_pi0}{\bf a}.  \new{This enhancement arises from an effect of the Cu $2p$ core-hole potential, namely, a shake-up of the photo-excited electron from the IGB to the UHB. As shown in Fig.~\ref{Fig.1}{\bf b},  the onset of the electron-hole pair excitation shares the same resonant energy of incident photons with the $dd$ excitation, because the $2p$ core-hole potential pulls down the UHB in the intermediate state of RIXS. Therefore,  hybridization between the IGB and the UHB is increased, inducing the shake-up from the IGB to the UHB.} In other words, the electron is first excited to the IGB near $(\pi, \pi)$ from Cu $2p$ and then further excited to the UHB through the shake-up process. A high-energy exciton is hence formed by an electron in the UHB and a hole in the coherent LHB near $(0, \pi)$. Although the TCFM is a low-energy effective model to describe the IGB and the coherent LHB and takes no account of the high-energy exciton involving the UHB, the same mechanism ascribed to the electron fractionalization near the Fermi level discussed above successfully accounts for the enhancement of the high-energy exciton in the same manner. 

In contrast, the spectral-weight transfer from high to low energies in the superconducting state relative to the normal phase was observed in the optical conductivity~\cite{molegraaf2002}, though a small amount. Such transfer, also suggested by the uniform mixing of the UHB and LHB in a simple model~\cite{phillips2020}, may imply a mechanism similar to the fractionalization here at work. However, suppose the UHB-LHB mixing uniformly enhances the LHB weight in the superconducting state; it is unclear why the enhancement is not observed at low energies around the elastic peak in the experimental RIXS spectra contrary to the naive expectation from this picture.

Interestingly, previous optical data ~\cite{Rubhausen2001} also show enhancement of the imaginary part of the dielectric constant in the order of 10\% for an underdoped sample for the energy around 3-3.5 eV. This result is indeed consistent with ours in terms of the fractionalization mechanism and further supports the universal feature of electron fractionalization, because the same mechanism of the enhancement in the superconducting phase works as the enhancement for LHB-UHB excitons observed around 2-3 eV in Figs.~\ref{OP_pi0}~and~\ref{O_K}, as is expected from the the optical transition from the coherent LHB at $(0,\pi)$ to the UHB at the same momentum that has substantial density of states (see Fig.~\ref{Fig.1}{\bf a}).

In conclusion, Cu $L_3$- and O $K$-edge RIXS spectra of OP Bi2212 consistently show an enhancement of the exciton intensity in the superconducting phase relative to the normal phase with a pseudogap. The enhancement is consistent with the prediction of the TCFM that electron fractionalization is suppressed in the superconducting phase.  The quasiparticle component dominant  above the pseudogap temperature $T^\ast$ partly recovers below $T_c$. We have discussed the enhancement without assuming any symmetry breaking because a simple mean-field-type symmetry breaking does not generate the dark fermion that is responsible for the enhancement. Therefore, spontaneous symmetry breaking such as charge order and nematicity proposed as mechanisms of the pseudogap would be difficult to explain the enhancement of RIXS intensity in the superconducting phase. In this sense, the present enhancement will pose severe constraints on the interpretation of the pseudogap.

The exciton peaks in the RIXS spectra show a substantial broadening. If intrinsic, it gives a rough estimate of the exciton lifetime in the order $\sim 0.1$ eV$^{-1}$. The possible interpretation of the dark fermion coupled to the exciton mentioned above offers a consistent picture, and the dark fermion lifetime is estimated in the same order. The large self-energies of excitons and dark fermions make it necessary to take into account the interaction of the dark fermion in future more elaborate studies. This enables us to disclose the nature of the dark fermion and its strong quantum entanglement on the microscopic level beyond the TCFM and  opens a possible route to studies on Planckian fluids~\cite{sachdev2010,zaanen2019,hartnoll2015} as suggested in the analysis of ARPES data~\cite{yamaji2021}. This is an intriguing future issue.

\begin{figure}[t!] 
\centering
\includegraphics[width=0.85\columnwidth]{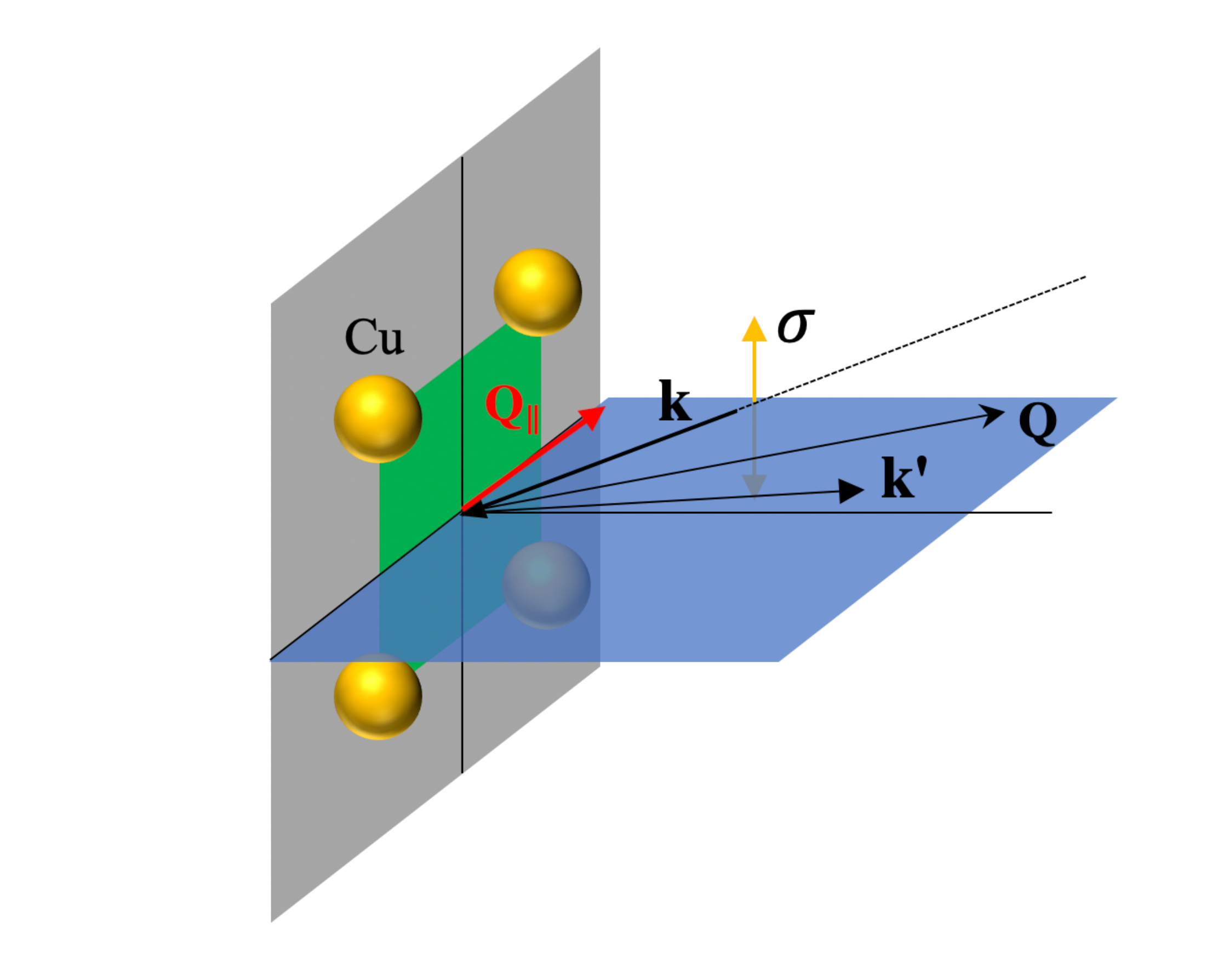}
\caption{\new{{\bf Scattering geometry of RIXS measurements}. The scattering plane is defined by the incident and scattered wave vectors, {\bf k} and $\bf k^\prime$, respectively. The projection  of the momentum transfer onto the CuO$_2$ plane, ${\bf Q}_{\|}$, is along the anti-nodal direction $(\pi, 0)$;  {\bf Q} is defined as $\bf k^{\prime}-${\bf k}.}}\label{fig_scattering}
\end{figure}

\vspace{10mm}
\noindent
{\bf Methods}
\vspace{1mm}

\new{Single crystals of pristine Bi$_{2.1}$Sr$_{1.9}$CaCu$_2$O$_{8+\delta}$ (Bi2212) and Pb-doped Bi$_{1.6}$Pb$_{0.4}$Sr$_2$CaCu$_2$O$_{8+\delta}$ (Pb-Bi2212) were grown in the air using the traveling solvent floating zone method \cite{watanabe1997}. The hole doping levels of the crystals were obtained using Tallon's empirical relation \cite{obertelli1992}. The crystals were then annealed to make them optimally-doped (OP) with $T_c=$ 89 K and 93~K for Bi2212 and Pb-Bi2212, respectively, under an oxygen partial pressure of 100 Pa at 600$^\circ$~C. The doping level of the OP crystals was  $p=0.16\pm0.005$. An over-doped Pb-Bi2212 crystal was also prepared with $T_c = 65$ K and $p=0.22\pm0.005$.}

We conducted Cu $L_3$-edge and O $K$-edge RIXS measurements at the AGM-AGS spectrometer of beamline 41A at Taiwan Photon Source \cite{singh2021}. XAS was recorded at normal incidence with $\sigma$ polarization using the total-electron-yield method. RIXS spectra across $T_c$ were recorded with $\sigma$ polarized incident X-rays of which the polarization was perpendicular to the scattering plane, as illustrated in Fig. \ref{fig_scattering}. The scattered X-rays were detected without a polarization analysis.

We calculated the exciton intensities in RIXS for the superconducting and pseudogap phases with ${\bf Q}_{\|} = (\pi, 0)$ and $(\pi/2, 0)$ using the TCFM, which is 
defined by the following Hamiltonian:
\begin{eqnarray}
H&=&\sum_{k,\sigma}[ \epsilon_c (k)c_{k,\sigma}^{\dagger}c_{k,\sigma} +\epsilon_d (k)d_{k,\sigma}^{\dagger}d_{k,\sigma}  \nonumber 
\\
&+& \Lambda (k) (c_{k,\sigma}^{\dagger}d_{k,\sigma} +{\rm H.c.})
\nonumber 
\\
&+&(\Delta_c(k) c_{k,\sigma}^{\dagger}c_{-k,-\sigma}^{\dagger}+\Delta_d(k) d_{k,\sigma}^{\dagger}d_{-k,-\sigma}^{\dagger} + {\rm H.c})
], \nonumber \\
\label{TCfermionAnomalous} 
\end{eqnarray}
where the fermion $c$ represents the original quasparticle with the dispersion $\epsilon_c(k)$ at the momentum $k$ in a form of a noninteracting Hamiltonian. The dark fermion  represented by $d$ with the dispersion $\epsilon_d(k)$ emerging from the strong correlation of the electrons hybridizes to the fermion $c$ via the coupling $\Lambda (k)$.
\new{We employ a square lattice with simple dispersion for $\epsilon_c (k)$ and $\epsilon_d (k)$ as
\begin{eqnarray}
\epsilon_c(k)=-(2t_{c1}(\cos k_x+\cos k_y)+4t_{c2}\cos k_x\cos k_y)+\mu_c, \nonumber \\
\epsilon_d(k)=-(2t_{d1}(\cos k_x+\cos k_y)+4t_{d2}\cos k_x\cos k_y)+\mu_d, \nonumber \\
\label{dispersion} 
\end{eqnarray} 
where $t_{c1}$ ($t_{d1}$) and $t_{c2}$ ($t_{d2}$) represent the nearest-neighbor and next-nearest-neighbor hoppings of $c$ ($d$) fermion, respectively.
The terms proportional to $\Delta_c(k)$ and $\Delta_d(k)$ represent the anomalous part in the mean-field approximation emerging in the superconducting state. Here we assume the simple $d$-wave superconducting gap 
\begin{eqnarray}
\Delta_c(k)=\frac{\Delta_{c0}}{2}(\cos k_x-\cos k_y), \nonumber \\
\Delta_d(k)=\frac{\Delta_{d0}}{2}(\cos k_x-\cos k_y).
\label{SCterm} 
\end{eqnarray} 
and the hybridization in the form
$\Lambda(k)=\Lambda_0+\Lambda_1(\cos k_x+1)(\cos k_y+1)$.
The parameter values are fitted to reproduce the ARPES~\cite{kondo2011} data together with its machine learning analysis~\cite{yamaji2021} and STM data~\cite{fischer2007} as  $t_{c1}=0.1953$, $t_{c2}=-0.0762$, $t_{d1}=0.0100$, $t_{d2}=-0.0036$, $\mu_c=0.2175$, $\mu_d=0.0105$, $\Delta_{c0}=0.02$, $\Delta_{d0}=0.07$, 
 $\Lambda_0=0.0658$
and  $\Lambda_1=-0.014$ in the unit of eV.} 
For the details of the TCFM calculations, see Supplementary Information and Refs.~\cite{imada2021,  imada2021rixs}. 

The RIXS intensity at the momentum ${\bf Q}$ and frequency $\omega$ is calculated using the formula
\begin{widetext}
\begin{eqnarray}
I_{\rm RIXS}({\bf Q},\omega,\omega_0;\sigma, \rho)&\propto&\sum_l|B_{li}({\bf Q}, \omega_0;\sigma,\rho)|^2(E_l-\omega+i\eta)^{-1}, 
\label{I_RIXS_1}
\end{eqnarray}
\begin{eqnarray}
B_{li}({\bf Q},\omega_0;\sigma,\rho) = \sum_{m,j}&&e^{i{\bf Q}\cdot {\bf R}_m}\chi_{\rho,\sigma}\langle l|c_{m\sigma}|j\rangle 
\langle j|(\omega_0-E_j+i\Gamma)^{-1}|j\rangle 
\langle j |c^{\dagger}_{m\rho}|i(k=0)\rangle,  
\label{I_RIXS_2} 
\nonumber\\
\end{eqnarray}
\end{widetext}
where $|i(k=0)\rangle$ is the ground state and $\chi_{\rho,\sigma}$ is the spin ($\rho$ and $\sigma$) dependent matrix representing  the product of two dipole matrix elements for absorption and emission. The core-hole lifetime width and exciton lifetime width are denoted by $\Gamma$ and $\eta$, respectively. 
The X-ray excites a core electron to the level $\omega_0$ measured from $E_{\rm F}$.
Here,  $c_m^{\dagger}$ is the local electron creation operator at the $m$-th site with the coordinate ${\bf R}_m$. 
The one-electron excited state represented by $|j\rangle$ is an eigenstate of the Hamiltonian (\ref{TCfermionAnomalous}) with energy $E_j$. All of them are relevant to the experimental setup of the crystal axes, X-ray polarization and momentum transfer.

\vspace{5mm}
\noindent\textbf{Data Availability}

All data generated or analysed during this study are available from the corresponding authors upon reasonable request.

\vspace{5mm}
\noindent\textbf{Acknowledgements}

This work was supported in part by the Ministry of Science and Technology of Taiwan under Grant No.~109-2112-M-213-010-MY3 and 109-2923-M-213-001.  
We also thank the support of KAKENHI Grant No.~15H02109,  No.~16H0634, No. 19K03741, No.~20K03849, and No.~22K03535 from JSPS. One of the authors, TW, was supported by a Hirosaki University Grant for Distinguished Researchers FY2017-2018. This research was also supported by MEXT as “Program for Promoting Researches on the Supercomputer Fugaku” (Basic Science for Emergence and Functionality in Quantum Matter - Innovative Strongly Correlated Electron Science by Integration of Fugaku and Frontier Experiments -, JPMXP1020200104) together with computational resources of supercomputer Fugaku provided by the RIKEN Center for Computational Science (Project ID: hp200132, hp210163 and hp220166).

\vspace{5mm}
\noindent\textbf{Author contributions}

D.J.H. and M.I. coordinated the experimental and theoretical works, respectively. A.S., H.Y.H., J.D.X., J.O., D.J.H., and C.T.C. conducted the RIXS experiments.  A.S., H.Y.H., J.D.X., D.J.H., and A.F. analyzed the data. M.I. performed the TCFM calculations. TW synthesized and characterized the Bi2212 samples. D.J.H., M.I., A.F., and A.S. wrote the manuscript with inputs from other authors.

\vspace{2mm}
\noindent\textbf{Competing interests}

The authors declare that there are no competing interests.

\bibliographystyle{naturemag}
\bibliography{ref_Bi2212_exciton_mi}

\end{document}



\baselineskip24pt


\maketitle


\begin{sciabstract}
\end{sciabstract}

\noindent {\bf This PDF file includes:}\\
\indent Electronic structure of cuprates\\
\indent Materials and Methods\\
\indent Figures S1 to S8\\
\indent References {(1-18)}

\newpage
\section*{Electronic structure of cuprates}

Figure S1 shows spectral function and dispersion of a hole-doped square-lattice calcuated from the Hubbard model. 
The hopping integral and the on-site Coulomb interaction used to simulate the cuprates are $t=0.5$~eV  and $U=4$ eV, respectively. Single-particle spectral function along a symmetric momentum line are reproduced from Fig.~7(a) of Ref.~\cite{charlebois2020}  at hole doping 12.5\%. The calculations were performed by a variational Monte-Carlo method. The coherent LHB and the incoherent LHB are denoted by ``coh LHB" and ``incoh LHB," respectively. Figure S1{\bf b} plots dispersions of the coherent LHB, the IGB, and a dark fermion band in the pseudogap energy region indicated by the blue dashed lines, adopted from Fig.~2(a) of Ref.~\cite{sakai2009}, where the cluster dynamical mean-field method was employed for hole doping 9\%. Occupied regions for an electron are highlighted in blue. The green surfaces plot the poles of the single-particle Green's function $G({\bf k}, \omega)$, showing peaks of single-particle spectral weights. The red surface plots the zeros of $G({\bf k}, \omega)$, which generates the pseudogap in this energy region and represents the dark fermion band in the present interpretation. The in-plane momenta ${\bf k}$ are given in units of $1/a$, in which $a$ is the lattice parameter.  

In the calculations shown in Fig.~S1{\bf b}, the pseudogap is characterized by the existence of the dark fermion band plotted as the red surface representing the zeros of the Green's function. For the physical picture of the zero of Green's function and the dark fermion dispersion to capture the essence of the pseudogap, readers are referred to Sec. {\bf Two-component fermion model}  and Ref.~\cite{imada2019}.

\newpage
\section*{Materials and Methods}
\subsection*{Samples}

Optimally-doped single crystals of Bi$_{2.1}$Sr$_{1.9}$CaCu$_2$O$_{8+\delta}$ (Bi2212) were grown in air using the traveling solvent floating zone method \cite{watanabe1997}. The crystals were then annealed  under an oxygen partial pressure of 100 Pa at 600$^\circ$ C to realize optimal doping with $T_c=$ 89 K. 
The hole doping level of $p=0.16\pm0.005$ was obtained using Tallon's empirical relation \cite{obertelli1992}. 
Figure~\ref{resistivity} shows that a weak pseudogap opens below $\sim$220 K according to the $c$-axis resistivity and a strong pseudogap open below $\sim$160 K according to the $ab$-plane resistivity \cite{usui2014}. Pb-doped single crystals of 
Bi$_{1.6}$Pb$_{0.4}$Sr$_2$CaCu$_2$O$_{8+\delta}$ were also prepared: Optimally-doped samples with $T_c = 93$ K and $p=0.16\pm0.005$, and overdoped ones with $T_c = 65$ K and $p=0.22\pm0.005$. For the overdoped samples, a weak pseudogap opens below $\sim$130 K according to the $c$-axis resistivity but the $ab$-plane resistivity shows no signature of a strong pseudogap \cite{usui2014}.

\subsection*{Cu $L$-edge  and O $K$-edge RIXS measurements}

We conducted Cu $L$-edge   and O $K$-edge RIXS measurements using the AGM-AGS spectrometer of beamline 41A at Taiwan Photon Source of National Synchrotron Radiation Research Center, Taiwan. This recently constructed AGM-AGS beamline is based on the energy compensation principle of grating dispersion; it has achieved an energy resolution of 16 meV at 530 eV photon energy \cite{singh2021}.  The instrument energy resolution  was 90~meV FWHM for  Cu $L$-edge RIXS measurements. 

We used RIXS to measure the excitonic excitation of Bi2212. Figure S3(a) shows a simplified graphic illustration of the excitation of a low-energy exciton in a hole-doped system induced by RIXS. The energy levels of all three stages of the RIXS excitation are aligned to $E_{\rm F}$. A Cu $2p$ electron is excited to the IGB with an energy $\omega_0$ above $E_{\rm F}$, and then another electron in the coherent LHB decays to fill the $2p$ core hole through emitting another photon of energy $\omega^\prime$. The UHB and incoherent LHB are pulled down by the $2p$ core hole in the intermediate state.

XAS spectra, as plotted in Fig. S3(b), were recorded at normal incidence with  $\sigma$ polarization by using the total electron yield method. RIXS spectra above and below $T_c$
were recorded with $\sigma$ polarized incident X-rays of which the polarization was perpendicular to the scattering plane.
Prior to RIXS measurements, a clean sample surface (001) was obtained on cleaving the sample in air.  
For the RIXS measurements, the scattering angle was fixed to 150$^\circ$. The $a$-axis and $c$-axis lay in the horizontal scattering plane while the $b$-axis was perpendicular to the scattering plane.  The scattered X-rays were detected without a polarization analysis. 

We extrapolated the fluorescence to a threshold, i.e., a zero energy loss, from the linear extrapolation of energy loss $E_{\rm loss}$ vs. incident photon energy. \new{The fluorescence threshold lead us to identify the energy position corresponding to the center of the occupied Cu $3d$ band; we found that it is 2~eV below the XAS absorption peak $L_3$. Because LDA calculations \cite{Krakauer1988} and resonant photoemission results \cite{Tjeng1992} indicate that the center of the occupied Cu $3d$ band is $\sim$3 eV below the Fermi level $E_{\rm F}$, the position of $E_{\rm F}$ in the absence of the core-hole potential is determined to be 1~eV above $L_3$. Our observation is consistent with the conclusions of the Cu $2p$ core-level photoemission \cite{sekhar1993} and absorption peak energies \cite{nucker1989}, which suggest that the Cu $L_3$ peak is located $\sim$1~eV below $E_{\rm F}$. 
Therefore, if we assume that relevant electronic states near $E_{\rm F}$ such as states in the IGB and coh LHB are shifted downwards by $\sim$1~eV owing to the core-hole potential, the energies of these electronic states relative to $E_{\rm F}$ are given by the energies relative to the $L_3$ absorption peak, as illustrated in Fig. S3(c). }

Supplementary XAS and RIXS data are presented in Figs. S4, S5, and S6, for O $K$-edge XAS spectrum of Bi-2212, Cu $L_3$-edge RIXS of optimally doped Bi2212 with ${\bf Q}_{\|} = (\pi/2, 0)$ and over-doped Pb-Bi2212 with ${\bf Q}_{\|} = (\pi, 0)$, respectively.

\subsection*{Two-component fermion model}

For the self-contained description, we summarize the structure of the Green's function and the self-energy of the two-component fermion model (TCFM) by referring to Refs.~\cite{yang_rice_zhang2006,sakai2016,imada2019}.
The TCFM in the normal state is defined by the following Hamiltonian:
\begin{eqnarray}
H&=&\sum_{k,\sigma,\sigma'}[ \epsilon_c (k)c_{k,\sigma}^{\dagger}c_{k,\sigma} +\Lambda (k) (c_{k,\sigma}^{\dagger}d_{k,\sigma} +{\rm H.c.}) \nonumber \\
&+&\epsilon_d (k)d_{k,\sigma}^{\dagger}d_{k,\sigma} ].
\label{TCfermionNormal} 
\end{eqnarray}
 Here, the fermion $c$ represents the original quasparticle with the dispersion $\epsilon_c(k)$ at the momentum $k$ in a form of a noninteracting Hamiltonian. The dark fermion  represented by $d$ with the dispersion $\epsilon_d(k)$ emerging from the strong correlation of the electrons hybridizes to the fermion $c$ via the coupling $\Lambda (k)$.

In the present interpretation of the doped cuprates, the original single-band system representing the antibonding band generated by strongly hybridized Cu $3d_{x^2-y^2}$ and O $2p_{\sigma}$ orbitals is under strong correlation. Then the interaction effect primarily yields emergent electron fractionalization, which results in the splintered two fermion degrees of freedom, $c$ and $d$ in the low-energy degrees of freedom near the fermi level. The fermion $c$, the normal quasiparticle component is built from the bare electron of antibonding band, while the strong correlation near the Mott insulator generates the effect beyond the conventional quasiparticle picture, which is better represented by the electron fractionalization. The emergent $d$ fermion represents such effects. The split of the degrees of freedom was, as one possibility, proposed to be generated from the electron bistability in the underdoped region. The bistability may also be the origin of the charge inhomogeneity or charge order widely observed in the cuprates~\cite{imada2021}. The fractionalization effect is represented in the original quasiparticle $c$ by the self-energy $\Sigma$, which we derive below. 
 
 Although Eq.(\ref{TCfermionNormal}) can be an effective phenomenological Hamiltonian of the symmetry broken phases such as stripe, nematic or time reversal symmetry broken states in the mean-field approximation, we should note that $c$ and $d$ are both visible particles by the spectroscopic measurements such as ARPES and the present RIXS in this case. This is because, for instance, $d_k$ is simply $c_{k+Q}$ in the translational symmetry broken phase with the ordering wave vector $Q$. In fact, such a symmetry broken phase does not yield the enhancement of the RIXS intensity we observed because the summation of detectable $c$ and $d$ density of states is preserved. The enhancement is purely a consequence of indetectabe nature of the $d$ fermion by the conventional spectroscopic measurements. The electron fractionalization, we assume, is then a consequence of NOT the symmetry breaking but simply the Mottness effect without the symmetry breaking.  The degrees of freedom $d$ is related to the broad background-like incoherent component in the spectroscopic measurements.
  The electron fractionalization gets increasing support as we summarize below.

The single-particle dynamics of the original quasiparticle $c$ in Eq.(\ref{TCfermionNormal}) is represented by the Green's function in the form of 
\begin{eqnarray}
G_c(k,\omega)=\frac{1}{\omega-\epsilon(k)-\Sigma (k,\omega)},
\label{Gkw} 
\end{eqnarray}
with the self-energy
\begin{eqnarray}
\Sigma(k,\omega)=\frac{\Lambda (k)^2}{\omega-\epsilon_d(k)}.
\label{TCMGkw} 
\end{eqnarray}
Equation (\ref{TCMGkw}) indicates that the pole of the self-energy emerges at $\omega= \epsilon_d(k)$, namely at the bare dispersion of the fermion $d$. This pole generates the zero of $G$ and a gap in the density of states of the fermion $c$ known as the hybridization gap, which is given as
\begin{eqnarray}
\Delta_{\rm HG}=\sqrt{(\epsilon_c(k)-\epsilon_d(k))^2+4\Lambda (k)^2}.
\label{HGap} 
\end{eqnarray}
In the TCFM picture, the pseudogap observed in the cuprates is well understood by this hybridization gap as is illustrated in Fig.~\ref{ARPES}(b).
Here the original visible quasiparticle $c$ is detected as the two split bands with the dispersion 
\begin{eqnarray}
\e_{\rm IGB}=\frac{1}{2}(\e_c+\e_d+\sqrt{(\e_c-\e_d)^2+4\L^2} \\
\e_{\rm cohLHB}=\frac{1}{2}(\e_c+\e_d-\sqrt{(\e_c-\e_d)^2+4\L^2}, 
\end{eqnarray}
where $\e_{\rm IGB}$ and $\e_{\rm cohLHB}$ represents the ingap band above the fermi level and the coherent LHB crossing the fermi level, respectively. 

To gain insight into the nature of the pseudogap formation, it is useful to examine the superconducting phase as well by introducing the $d$-wave superconducting mean field
as
\begin{eqnarray}
H&=&\sum_{k,\sigma}[ \epsilon_c (k)c_{k,\sigma}^{\dagger}c_{k,\sigma} +\epsilon_d (k)d_{k,\sigma}^{\dagger}d_{k,\sigma}  \nonumber 
\\
&+& \Lambda (k) (c_{k,\sigma}^{\dagger}d_{k,\sigma} +{\rm H.c.})
\nonumber 
\\
&+&(\Delta_c(k) c_{k,\sigma}^{\dagger}c_{-k,-\sigma}^{\dagger}+\Delta_d(k) d_{k,\sigma}^{\dagger}d_{-k,-\sigma}^{\dagger} + {\rm H.c})
], \nonumber \\
\label{TCfermionAnomalous} 
\end{eqnarray}
where the anomalous part proportional to the superconducting order parameters $\Delta_c(k)$ and $\Delta_d(k)$ becomes nonzero. This is nothing but Eq.(1) in the main text.


Then Green's function for $c$ particle in Nambu representation is obtained as
\begin{eqnarray}
G_c(k,\omega)=\frac{1}{\omega-\epsilon_c(k)-\Sigma^{\rm nor}(k,\omega)-W(k,\omega)},
\label{SCGkw} 
\end{eqnarray}
with 
\begin{eqnarray}
W(k,\omega)=\frac{\Sigma^{\rm ano}(k,\omega)^2}{\omega+\epsilon_c(k)+\Sigma^{\rm nor}(k,-\omega)^*},
\label{SCGkw2} 
\end{eqnarray}
\begin{eqnarray}
\Sigma^{\rm nor}(k,\omega)=\frac{\Lambda(k)^2(\omega+\epsilon_d(k))}{\omega^2-\epsilon_d(k)^2-\Delta_d(k)^2},
\label{SCGkw3} 
\end{eqnarray}
and 
\begin{eqnarray}
\Sigma^{\rm ano}(k,\omega)=\Delta_c(k)-\frac{\Lambda(k)^2\Delta_d(k)}{\omega^2-\epsilon_d(k)^2-\Delta_d(k)^2}.
\label{SCGkw4} 
\end{eqnarray}

Now the pole position of $\Sigma^{\rm nor}$ at $\omega=\epsilon_d(k)$ in the normal state (expected to generate the pseudogap) is modified to  $\omega=\pm\sqrt{\epsilon_d(k)^2+\Delta_d(k)^2}$.
Remarkably, the anomalous part $\Sigma^{\rm ano}$ also has a pole exactly at the same position.  Accordingly, $W$ (Eq.(\ref{SCGkw2})) in the denominator of $G$ in Eq.(\ref{SCGkw}) also has a pole of the order 1 at the same energy. The residue of the poles of $W$ and $\Sigma^{\rm nor}$ at $\omega=\pm\sqrt{\epsilon_d(k)^2+\Delta_d(k)^2}$ are given by 
$\frac{\L^2}{2}\left(1\pm \frac{\e_d}{\sqrt{\e_d^2+\D_d^2}} \right)$ and $-\frac{\L^2}{2}\left(1\pm \frac{\e_d}{\sqrt{\e_d^2+\D_d^2}} \right)$, respectively. Therefore, their residues cancel out in their sum in Eq.~(\ref{SCGkw}) and will be hidden in the direct data of the spectral function measured by ARPES~\cite{sakai2016, imada2019}.
If the origin of the pseudogap is not ascribed to this fractionalization mechanism, such a remarkable cancellation would not be expected. The hidden pole structure was extracted by machine learning of ARPES data in Ref.~\cite{yamaji2021}.

In the present analysis, as are described in Eqs.(2) and (3) in the main text, we have employed the parameters of the TCFM~\cite{imada2021rixs} as follows:
 \begin{eqnarray}
\epsilon_c(k)=-(2t_{c1}(\cos k_x+\cos k_y)+4t_{c2}\cos k_x\cos k_y)+\mu_c, \nonumber \\
\epsilon_d(k)=-(2t_{d1}(\cos k_x+\cos k_y)+4t_{d2}\cos k_x\cos k_y)+\mu_d, \nonumber \\
\Lambda(k)=\Lambda_0+\Lambda_1(\cos k_x+1)(\cos k_y+1),
\label{dispersion} 
\end{eqnarray} 
 \begin{eqnarray}
\Delta_c(k)=\frac{\Delta_{c0}}{2}(\cos k_x-\cos k_y), \nonumber \\
\Delta_d(k)=\frac{\Delta_{d0}}{2}(\cos k_x-\cos k_y),
\label{SCterm} 
\end{eqnarray} 
where the parameters are chosen to reproduce the ARPES data\cite{kondo2011} analyzed by machine learning~\cite{yamaji2021} and STM data~\cite{fischer2007} as $t_{c1}=0.1953$, $t_{c2}=-0.0762$, $t_{d1}=0.0100$, $t_{d2}=-0.0036$, $\mu_c=0.2175$, $\mu_d=0.0105$, $\Delta_{c0}=0.02$, $\Delta_{d0}=0.07$, 
 $\Lambda_0=0.0658$
and  $\Lambda_1=-0.014$ in the unit of eV. 
In fact, these parameters well fit the ARPES data~\cite{kondo2011} and the machine learning analysis~\cite{yamaji2021}.
For instance the well-known peak-dip hump structure of the spectral weight ${\rm Im} G(k,\omega)$ for Bi2212 in the superconducting phase is quantitatively reproduced as is shown in Fig.~\ref{ARPES}. Here the quasiparticle peak energy is $\pm\sim 0.03$ eV and the hump energy at the antinodal point is around $\pm 0.12$ ev in agreement with the ARPES data~\cite{kondo2011}. Furthermore, the particle-hole  asymmetric structure of the density of states observed in scanning tunneling microscope~\cite{fischer2007} as well as the pseudogap size above $T_{\rm c}$ larger than the superconducting gap below $T_{\rm c}$ is also reproduced.

For the cancellation of the normal and anomalous component of the self-energy in the spectral function as is indeed the case of the machine learning result of the ARPES measurement, we are forced to employ the electron fractionalization. However, the ARPES data contains several limitations such as the range limit of detectable momentum and energy windows, uncertainty arising from unknown background effects and experimental noise. Therefore, it is desired to perform independent measurement by RIXS to stringently test the radical concept of the fractionalization, because of its fundamental importance. 

\noindent 

\bibliographystyle{naturemag}
\bibliography{ref_Bi2212_exciton_mi.bib}

\clearpage

\begin{figure}[t!]
\centering\includegraphics[width=0.95\columnwidth]{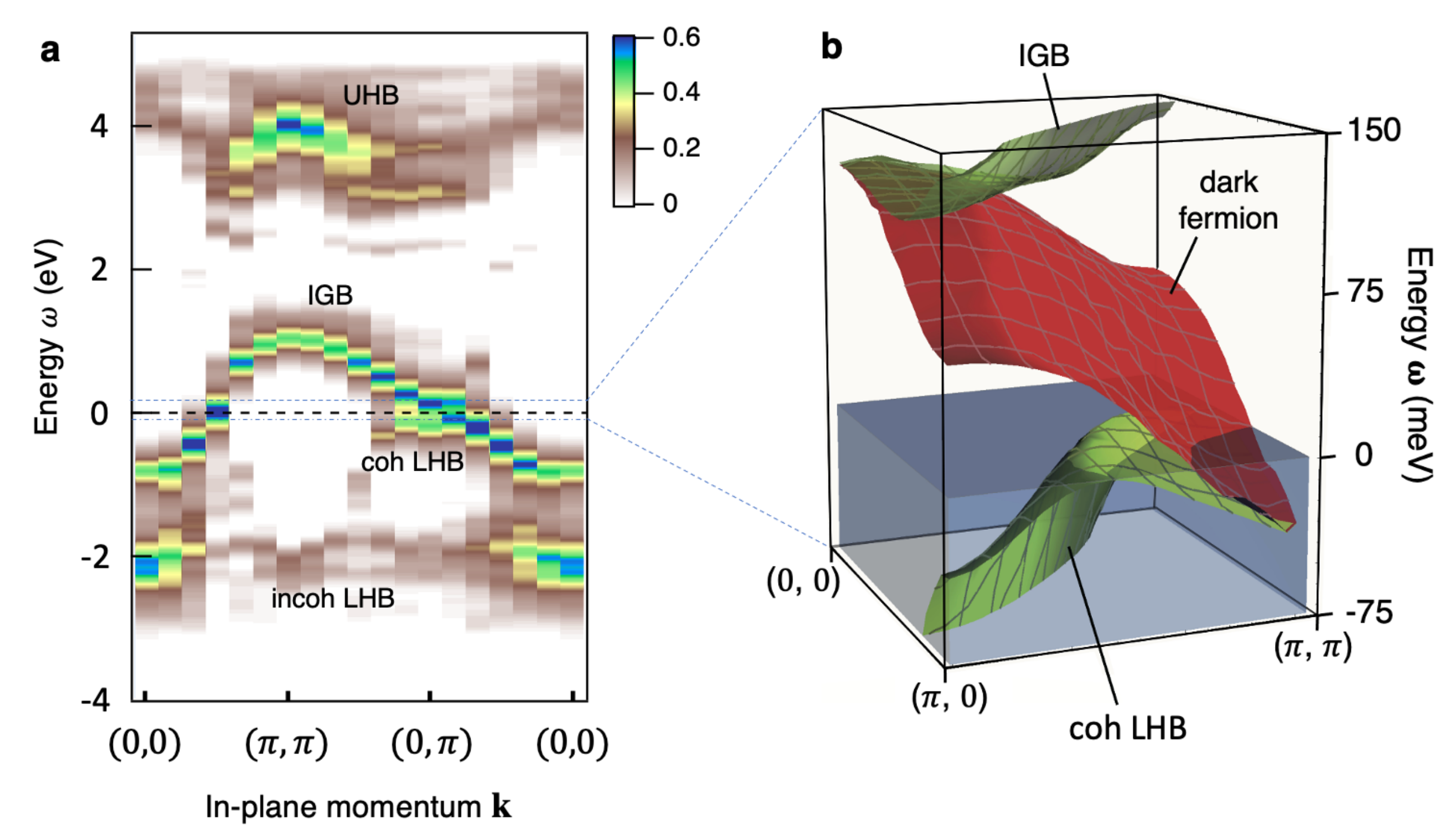}
\caption{{\bf a} Spectral function and dispersion of a hole-doped square-lattice Hubbard model. {\bf b} Dispersions of the coherent LHB, the IGB, and a dark fermion band in the pseudogap energy region.} 
\label{fig_S_darkfermion}
\end{figure}

\begin{figure*}[h]
\centering 
\includegraphics[width=\columnwidth]{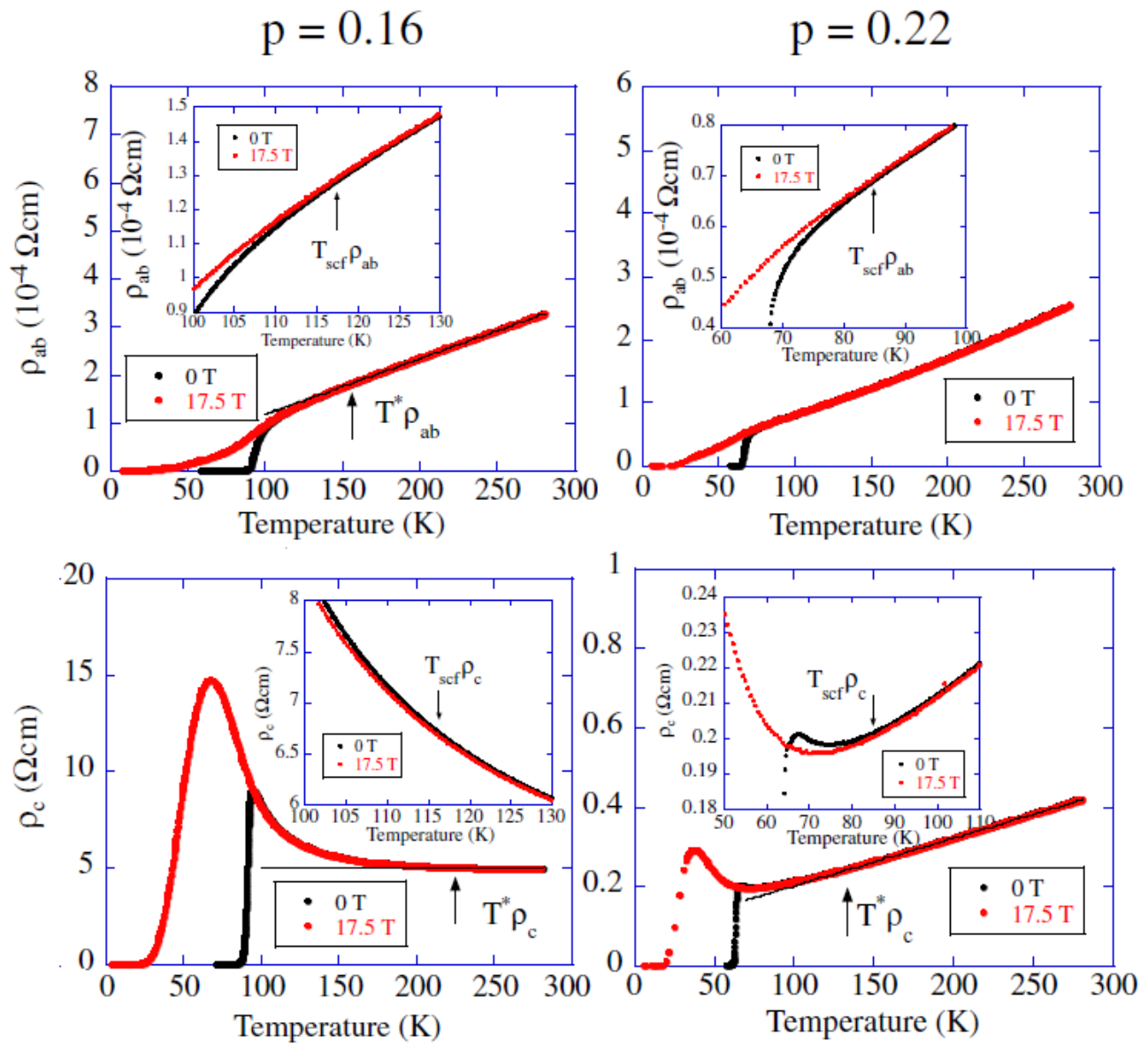} 
\caption{In-plane and out-of-plane resistivities $\rho_{ab}$ and $\rho_{c}$ of Bi2212 single crystals \cite{usui2014}. The temperatures $T^{\ast}_{\rho_{ab}}$ and  $T^{\ast}_{\rho_{c}}$ at which the resistivity starts to deviate
from the higher-temperature linear behaviors indicates the opening of pseudogaps. Left: Optimally-doped 
Bi$_{2.1}$Sr$_{1.9}$CaCu$_2$O$_{8+\delta}$ with $p=$ 0.16.  Right: Overdoped Bi$_{1.6}$Pb$_{0.4}$Sr$_{2}$CaCu$_2$O$_{8+\delta}$ with $p=$ 0.22. No signature of the pseudogap opening is seen in the $\rho_{ab}$ of the overdoped sample. In the inset, arrows indicate the temperatures $T_{\rm{scf}\rho_{ab}}$ and $T_{\rm{scf}\rho_{c}}$ below which superconducting fluctuations are significant and are suppressed by the magnetic field of 17.5 T.} 
\label{resistivity}
\end{figure*}

\begin{figure*}[h]
\centering 
\includegraphics[width=15cm]{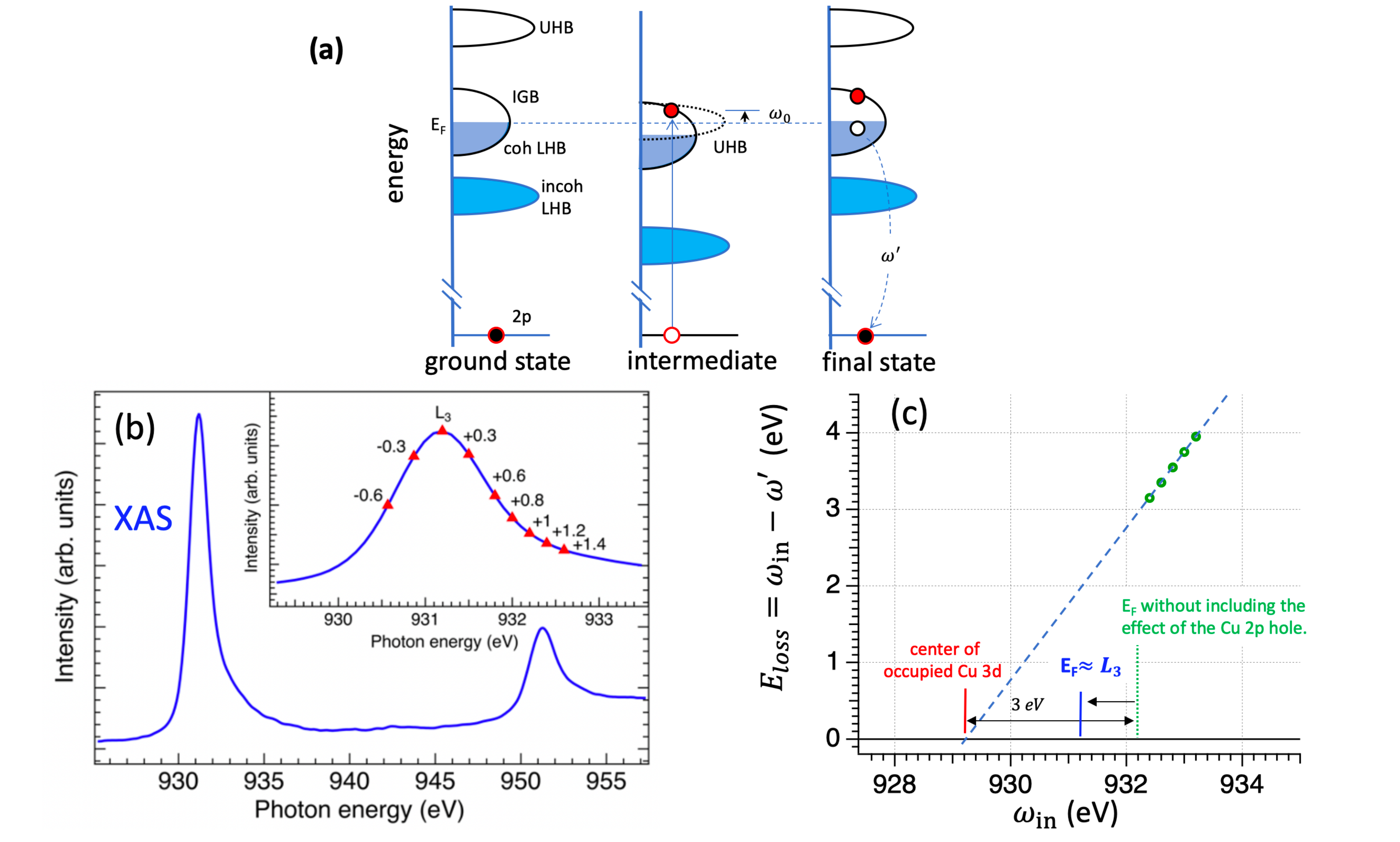}
\caption{ (a)~Simplified graphic illustration of the excitation of a low-energy exciton in a hole-doped system induced by RIXS. The energy levels of all three stages of the RIXS excitation are aligned to $E_{\rm F}$.  (b)~The Cu $L$-edge XAS spectrum of Bi-2212 measured with $\sigma$ polarization by using the total electron yield method. Inset: zoom-in XAS spectrum around $L_3$-edge; the peak energy is denoted as $L_3$. Red triangles indicate the energies used for the RIXS measurements. (c)~Estimate the energy position of $E_{\rm F}$ in the Cu $L_3$-edge absorption spectrum through the fluorescence threshold from the linear extrapolation of energy loss $E_{\rm loss}$ vs incident photon energy $\omega_{\rm in}$.} 
\label{Cu_XAS}
\end{figure*}

\begin{figure*}[h]
\centering 
\includegraphics[width=12cm]{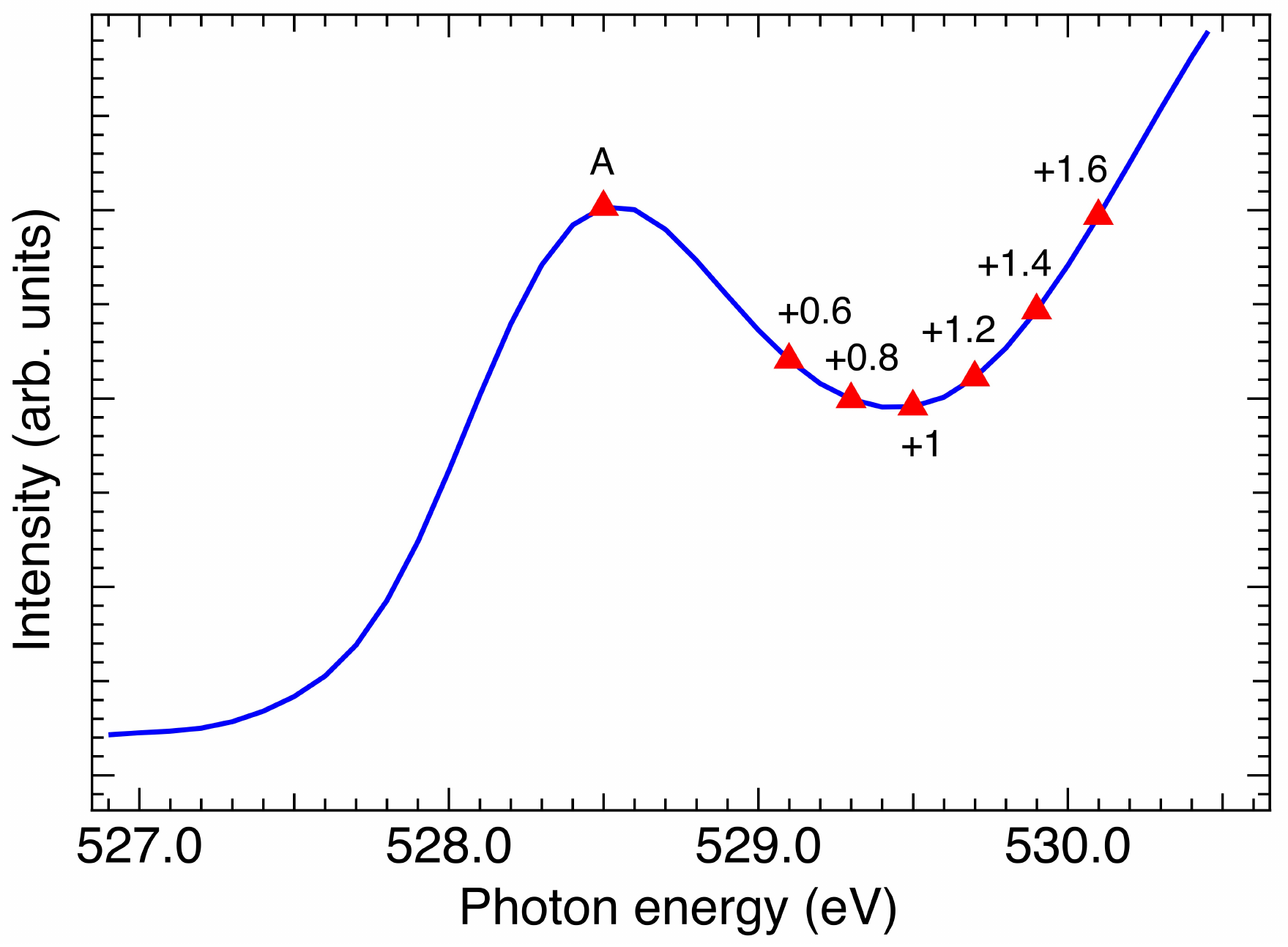}
\caption{ The O $K$-edge XAS spectrum of Bi-2212 measured with $\sigma$ polarization by using the total electron yield method. The resonant energy is denoted as A. Red triangles indicate the energies used for the RIXS measurements.} 
\label{O_XAS}
\end{figure*}

\begin{figure*}[t!]
\centering 
\centering 
\includegraphics[width=1\columnwidth]{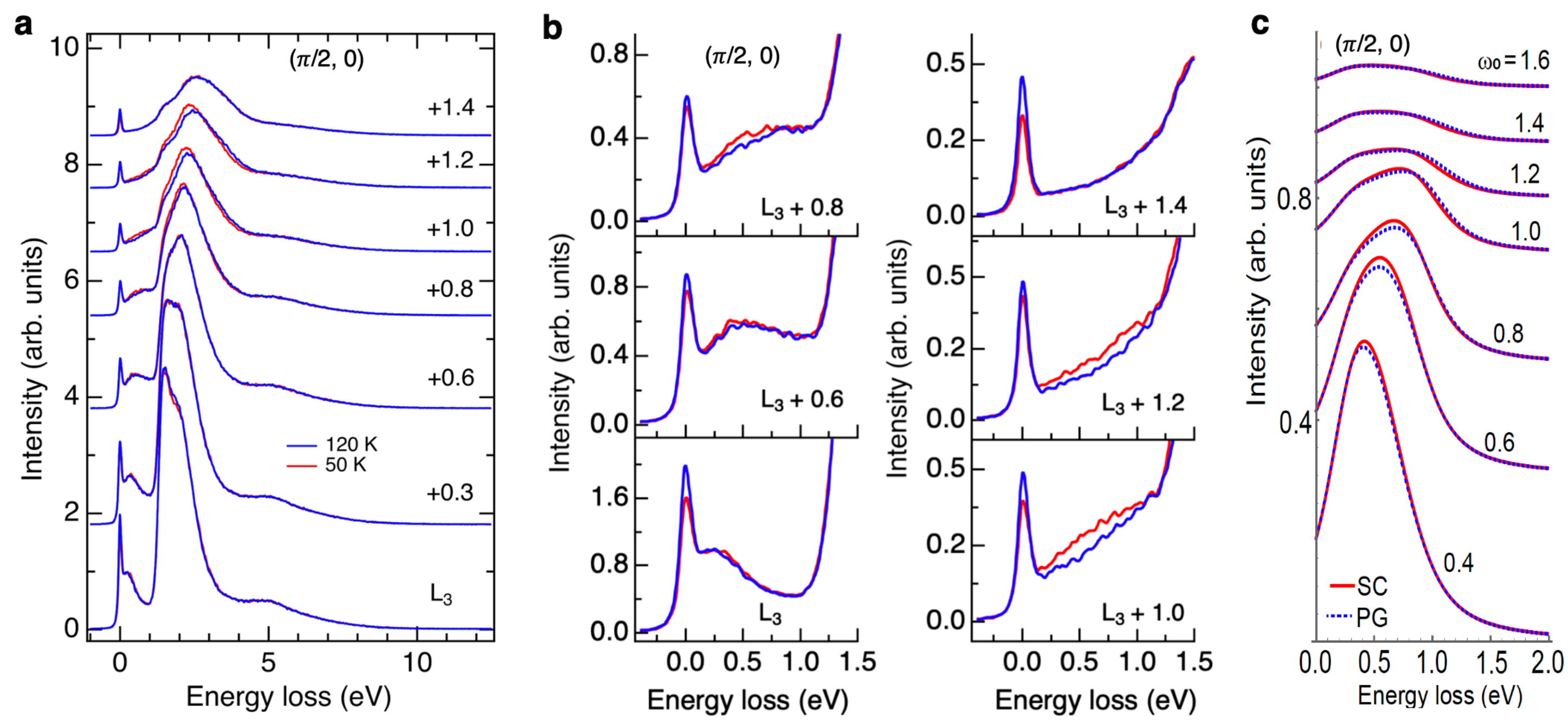} 
\caption{{\bf Enhancement of excitonic excitations in Cu $L_3$-edge RIXS of optimally doped Bi2212 by superconductivity}. {\bf a} \& {\bf b}~RIXS spectra at selected incident photon energies for ${\bf Q}_{\|} = (\pi/2, 0) $  and temperatures above and below $T_c$ = 89~K. The incident photon energy is denoted as its energy above the $L_3$ absorption in units of eV. Spectra above and below $T_c$ are normalized for energy loss from 1.7 eV to 13 eV. From the estimate shown in supplementary  Fig. S2, $L_3$  corresponds approximately to $E_{\rm F}$. {\bf c}~Calculated RIXS resulting from excitonic excitations in the superconducting (SC) and pseudogap (PG) phases for ${\bf Q}_{\|} = (\pi/2, 0)$. 
The incident energy $ \omega_0$ measured from $E_{\rm F}$ is given in units of eV.  The core-hole lifetime width $\Gamma$ and the broadening factor $\eta$ were set to 0.3~eV and 0.1~eV, respectively. All spectra are plotted with a vertical offset for clarity.} 
\label{fig_S_halfpi}
\end{figure*}

\begin{figure*}[h]
\centering 
\includegraphics[width=15cm]{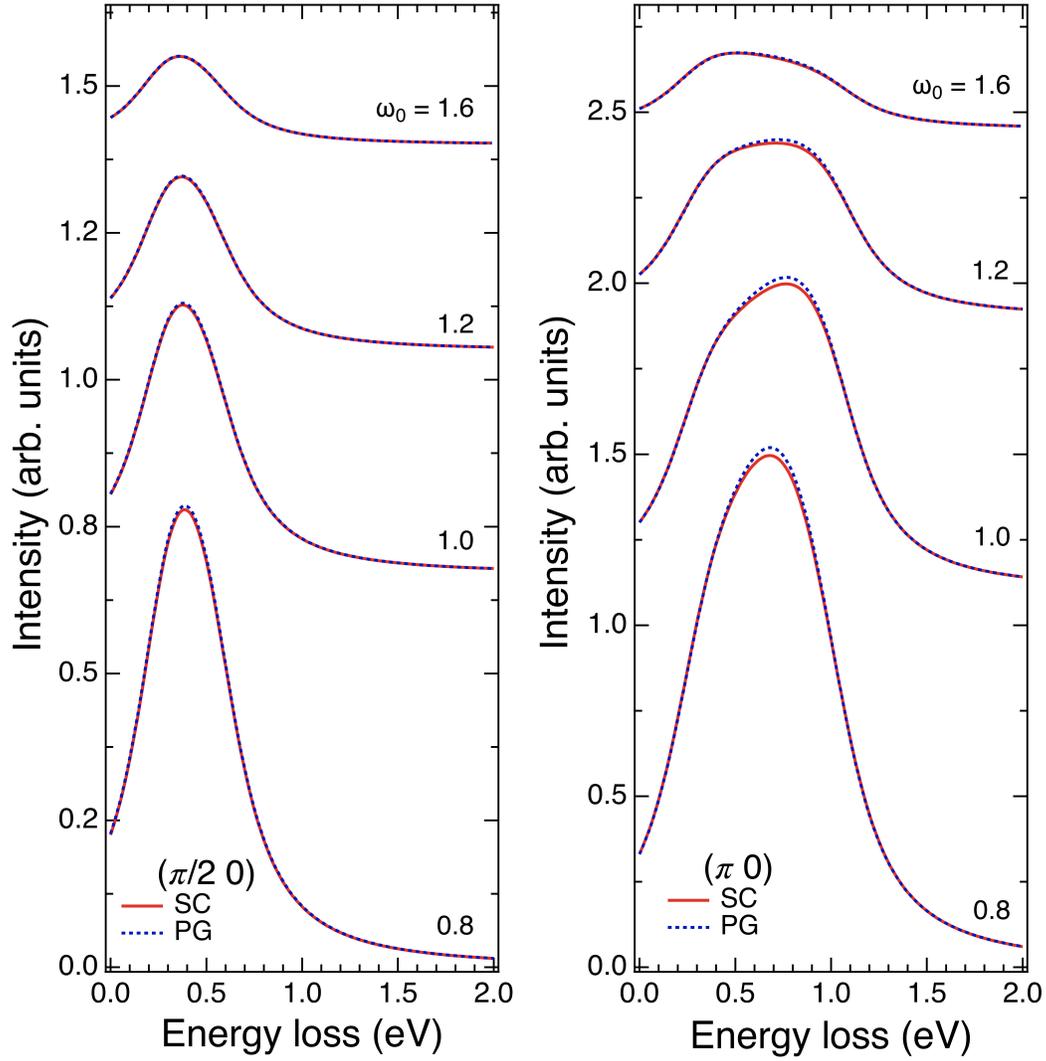}\caption{Calculated RIXS spectra using a conventional single-component Hubbard model for SC and PG phases at selected $\omega_{0}$ for ${\bf Q}_{\|} = (\pi/2, 0)$ and $(\pi, 0)$. All spectra are plotted with a vertical offset for clarity.} 
\label{fig_S_singleband}
\end{figure*}

\begin{figure*}[t!]
\centering 
\centering 
\includegraphics[width=12cm]{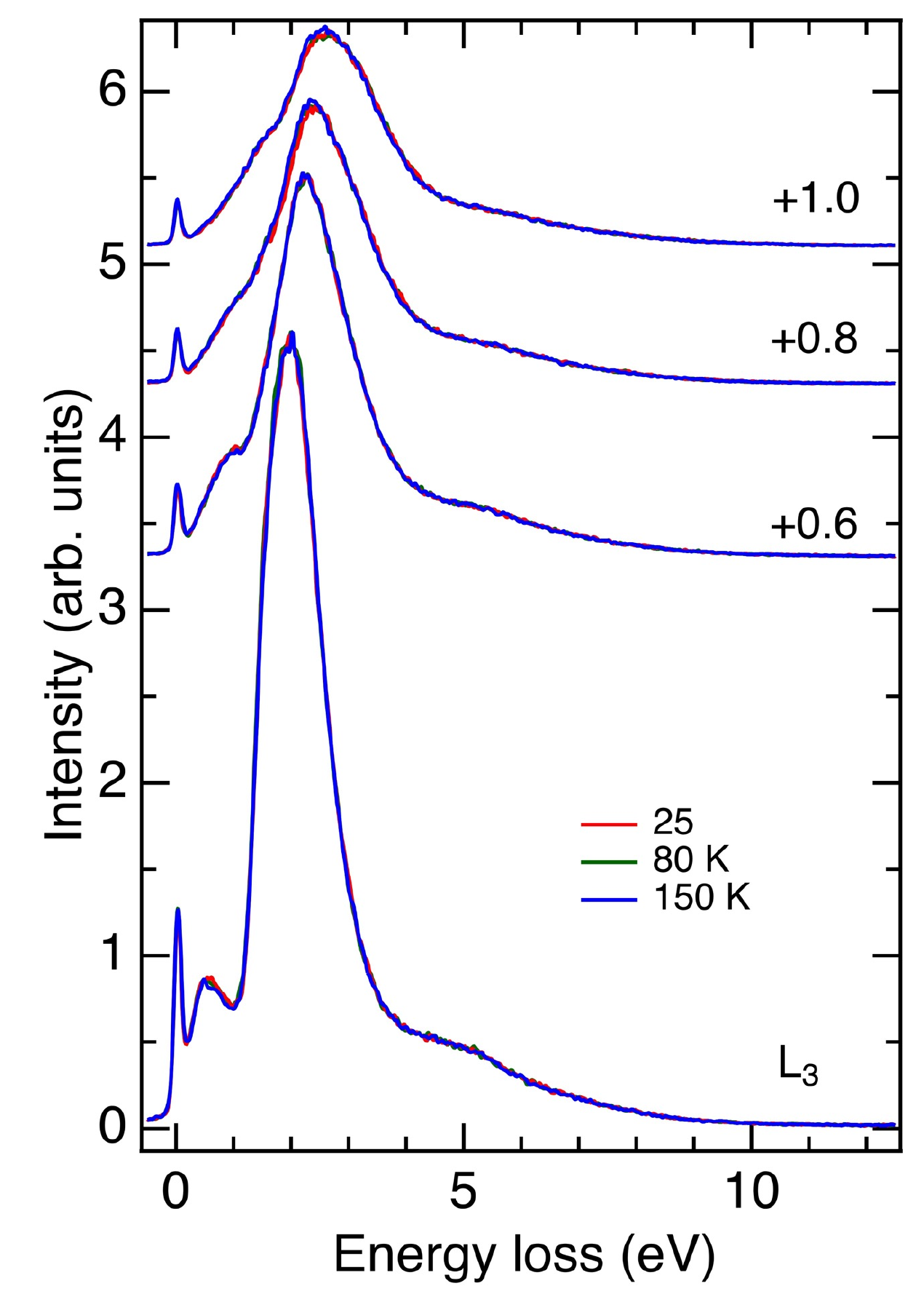} 
\caption{Cu $L_3$-edge RIXS spectra of OD Pb-Bi2212 measured at ${\bf Q}_{\|} = (\pi, 0)$ for various incident energies across the $L_3$ peak of the XAS. RIXS spectra were measured at 25K, 80K and 150K as plotted by red green and blue solid lines. Spectra are plotted with a vertical offset for clarity.}
\label{fig_S_OD}
\end{figure*} 

\begin{figure}[h]
\begin{center} 
\includegraphics[width=18cm]{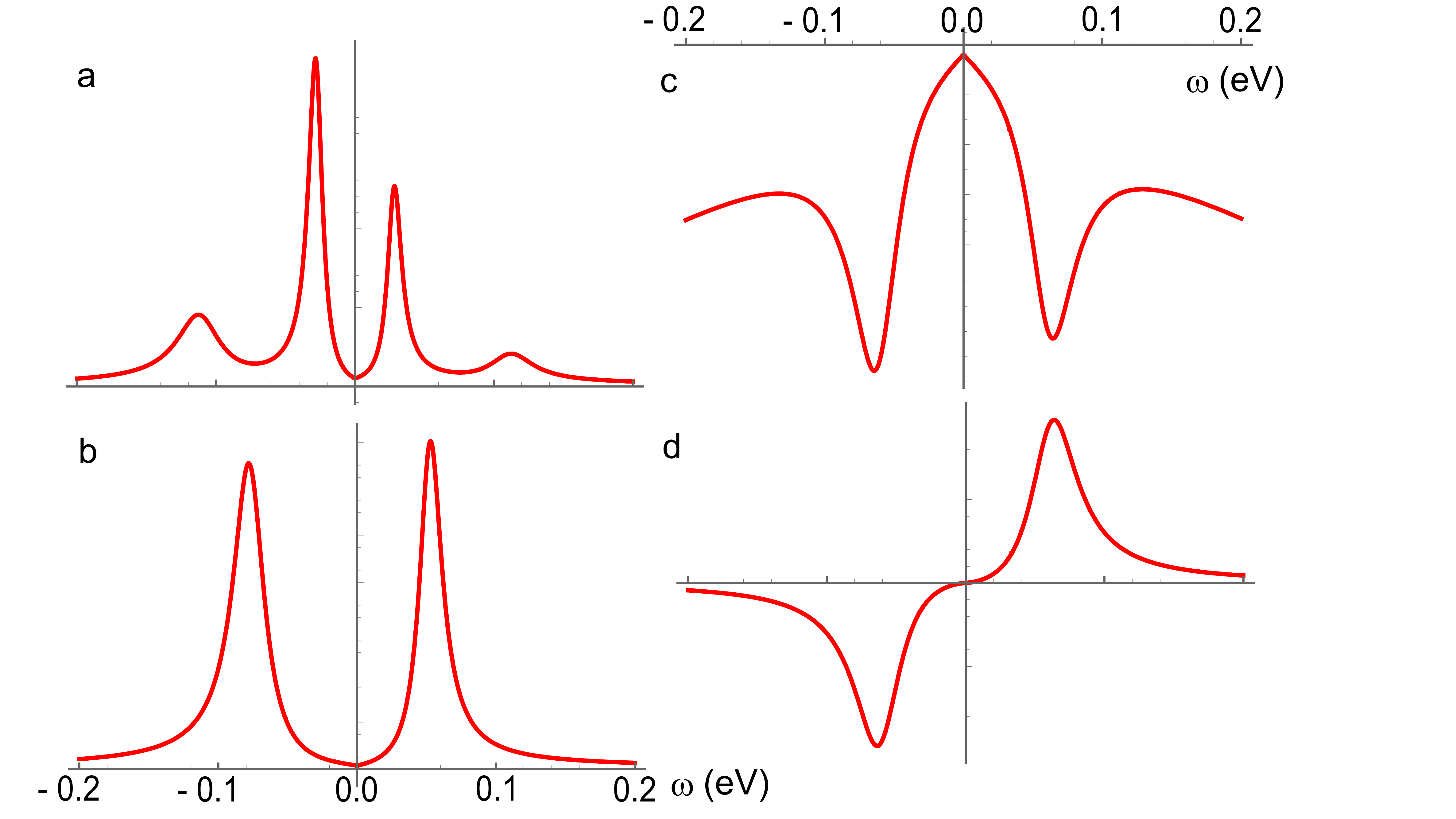}
\caption{{\bf a} Spectral function $A(k,\omega)$ obtained from TCFM in the superconducting phase to reproduce the ARPES~\cite{kondo2011} and the machine learning data~\cite{yamaji2021}. 
{\bf b} $A(k,\omega)$ in the normal (pseudogap) state for the same parameters as {\bf a} except for the choice  $\Delta_{d0}=\Delta_{c0}=0$. 
{\bf c}. Imaginary part of the normal self-energy for the superconducting state. {\bf d} Imaginary part of the anomalous self-energy for the superconducting state. In the superconducting state normal and anomalous contributions of the peaks to $A(k,\omega)$ cancels.
}
\label{ARPES}
\end{center}
\label{fig_S_Akw}
\end{figure}